\documentclass[12pt,notitlepage,a4paper]{article}

\pdfoutput=1

\usepackage{a4wide}
\usepackage{epsfig}
\usepackage{color,graphicx}
\usepackage{cite}
\usepackage{amssymb,amsmath}
\usepackage{changepage}
\usepackage{multirow}
\usepackage{braket}
\usepackage{youngtab}

\newcommand{\be}{\begin{equation}}
\newcommand{\ee}{\end{equation}}
\newcommand{\bea}{\begin{eqnarray}}
\newcommand{\eea}{\end{eqnarray}}

\begin{document}

\begin{center}  

\vskip 2cm 

\centerline{\Large {\bf Instanton operators and symmetry enhancement}}
\vskip 0.5cm
\centerline{\Large {\bf in $5d$ supersymmetric $USp$, $SO$ and exceptional gauge theories}}
\vskip 1cm

\renewcommand{\thefootnote}{\fnsymbol{footnote}}

   \centerline{
    {\large \bf Gabi Zafrir${}^{a}$} \footnote{gabizaf@techunix.technion.ac.il}}

\vspace{1cm}
\centerline{{\it ${}^a$ Department of Physics, Technion, Israel Institute of Technology}} \centerline{{\it Haifa, 32000, Israel}}
\vspace{1cm}

\end{center}

\vskip 0.3 cm

\setcounter{footnote}{0}
\renewcommand{\thefootnote}{\arabic{footnote}}   
   
      \begin{abstract}
     
		We study the fermionic zero modes around $1$ instanton operators for $5d$ supersymmetric gauge theories of type $USp$, $SO$ and the exceptional groups. The major motivation is to try to understand the global symmetry enhancement pattern in these theories. 
		
      \end{abstract}

\tableofcontents

\section{Introduction}

Gauge theories in $5d$ are non-renormalizable, and so naively are not good microscopic theories. However, at least for minimally supersymmetric $5d$ gauge theories, it is known that a fixed point can exist so these do in fact define a microscopic theory\cite{SEI,SM,SMI}. These theories in turn sometimes exhibit a peculiar phenomenon of symmetry enhancement in which the fixed point exhibit a larger global symmetry than the gauge theory. An important element in this is the existence of a topologically conserved $U(1)$ current, $j_T = *\mbox{Tr}(F\wedge F)$, associated with every non-abelian gauge group. The particles, charged under this current, are instantons which are particles in $5d$. 

In many cases these instantonic particles provide additional conserved currents so that the global symmetry of the fixed point is larger than that of the gauge theory. A classic example is an $SU(2)$ gauge theory with $N_f$ hypermultiplets in the doublet of $SU(2)$. This theory is known to flow to a $5d$ fixed point provided $N_f\leq 7$\cite{SEI}. Furthermore, the global symmetry at the fixed point is enhanced from $U(1)\times SO(2N)$ to $E_{N_f+1}$\cite{SEI}. From the gauge theory viewpoint the enhancement is brought about by instantonic particles. 

The $N_f=8$ case is also interesting. In that case there is no $5d$ fixed point, but rather a $6d$ one\cite{GMS}. This theory also has an enhanced symmetry but this time the Lorentz symmetry is enhanced. Again, the additional conserved currents are brought about by instanton particles. A similar story is also believed to occur for the maximally supersymmetric theory, where the $6d$ theory is now the $(2,0)$ theory\cite{Dou,LPS}.

A natural question then is how can we determine if there is an enhancement and if so what is the enhanced symmetry. One way is to infer this from the brane web presentation of the theory\cite{HA,HAK,DHIK}. A more direct way is to find this from the superconformal index\cite{KMMS}. The superconformal index is a counting of the BPS operators of the theory where the counting is such that if two operators can merge to form a non-BPS multiplet they will sum to zero. Particularly for 5d SCFT's, the theory is considered on $S^{4} \times S^1$. Then the representations of the superconformal group are labeled by the highest weight of its $SO_L(5) \times SU_R(2)$ subgroup. We will call the two weights of $SO_L(5)$ as $j_1, j_2$ and those of $SU_R(2)$ as $R$. Then following \cite{KKL} the index is:

\be
\mathcal{I}={\rm Tr}\,(-1)^F\,x^{2\,(j_1+R)}\,y^{2\,j_2}\,\mathfrak{q}^{\mathfrak{Q}}\,. \label{eq:ind}
\ee 
where $x,\,y$ are the fugacities associated with the superconformal group, while the fugacities collectively denoted by $\mathfrak{q}$ correspond to other commuting charges $\mathfrak{Q}$, generally flavor and topological symmetries. 

More to the point, conserved currents are part of a BPS supermultiplet which contains a scalar operator with charges $j_1=j_2=0, R=1$, and in the Adjoint of the corresponding global symmetry. Thus, it contributes to the index a term of $x^2 \chi[\bold{Ad}]$. Therefore one can calculate the index, and from the $x^2$ terms infer the expected enhancement.

There are two broad methods to calculate the index. One, directly from the gauge theory, relying on the localization calculation in \cite{KKL}. For examples of calculations using these methods, see \cite{KKL,Bash,BGZ,BGZ1,BGZ2,HKKP,Zaf,BZ}. Alternatively, given a brane web presentation, it can be calculated using methods of topological strings, as first done in \cite{IV}. For examples of calculations using these methods, see \cite{IV,BPTY,BMPTY,HKT,Taki,AHS,Taki1,HZ,HTY,MPTY}.

 While all of these methods are useful they have their shortcomings. First, they are very technical and in many cases of interest involve removing spurious contributions that need to be determined from additional input, for example using brane webs. Also, there are cases where these methods cannot be applied, for example, exceptional groups where both a brane web description and instanton counting are unknown. 

Recently, a simpler method for determining conserved currents coming from the $1$ instanton was proposed in \cite{Tachi}. Besides being simpler, it also appear to be applicable to any gauge group and any matter content. The original article dealt with gauge group $SU(N)$. The purpose of this article is to explore the results of this method to other gauge groups. We consider the simplest possible case, being a simple gauge group with matter in some representation of that group. A first question is what sort of matter representations should we allow. We want to look only at cases that flow to $5d$ or $6d$ fixed point, but it is not clear under what conditions a $5d$ gauge theory flows to such a fixed point. 

We adopt the following rather broad condition. It is widely believed that the maximally supersymmetric gauge theory, as well as $SU(2)+8F$, flow to a $6d$ fixed point. Also it is believed that adding additional flavors, of any non-trivial representation, leads to no fixed point. Thus, if a theory has a Higgs branch leading to such theories, it is reasonable that it has neither a $5d$ nor a $6d$ fixed point. We explore all cases where there is a Higgs branch leading to a pure gauge theory, maximally supersymmetric gauge theory or $SU(2)+N_fF$ with $N_f\leq 8$. This of course does not prove that these theories have a $6d$ or $5d$ fixed point. Since every $\mathcal{N}$$=2$ conformal $4d$ gauge theory obeys this condition, and as the Higgs branch is the same as in $5d$, one can largely borrow the classification of \cite{BT}, adjusting the cases flowing to $SU(2)+4F$ to allow additional flavors. 

The structure of this paper is as follows. We start in section $2$ with a review of the method. After that we move on to discus gauge groups $USp(2N)$ and $SO(N)$ in sections $3$ and $4$ respectively. Section $5$ covers the exeptional groups. We end in section $6$ with our conclusions.

\medskip

{\bf A word on notation:} We will denote global symmetries associated with 
matter in the fundamental representation (``flavor") by an $F$ subscript, those associated with
matter in the $2$-index antisymmetric representation by an $AS$ subscript, and those associated with
matter in the $3$-index antisymmetric representation by a $TAS$ subscript.

When discussing $SO(N)$ groups with spinor matter we will denote the spinor flavor symmetry by an $S$ subscript. When $N$ is dividable by $4$, there are two different, self-conjugate, spinor representations and we use also a $C$ subscript for flavor symmetry associated with the other type of spinors. 

In addition, we will use a $T$ subscript for the topological (instanton) $U(1)$ symmetries. If no subscript is written then this $U(1)$ is a global one that is a combination of the various $U(1)$'s in the theory. Subscripts on gauge symmetries will sometimes be used to denote the discrete $\theta$ parameter.

When denoting the charges of states under the $SU_R(2)$ R-symmetry, we use $R$ for the maximal weight of an $SU_R(2)$ multiplet, and $r$ for the $U_R(1) \subset SU_R(2)$ charge of a specific state in that multiplet.

\section{Review of the method}

We consider a $1$ instanton of an $SU(2)$ gauge theory. As the $1$ instanton breaks part of the spacetime and gauge symmetry, there are zero modes associated with these broken symmetries. This builds an $8$ dimensional moduli space. Besides these, when fermionic matter is present, there are also fermionic zero modes whose number depends on the matter representation. Specifically, a Weyl fermion in the doublet of $SU(2)$ gives a single fermionic zero mode while a Weyl fermion in the Adjoint of $SU(2)$ gives $4$ zero modes. As we consider supersymmetric theories, we always have gauginos that are in the Adjoint of $SU(2)$. These lead to $8$ zero modes for the $\mathcal{N}$$=1$ case and $16$ for the $\mathcal{N}$$=2$ case\footnote{The instanton configuration is BPS, and breaks half of the superconformal symmetry. The resulting fermionic zero modes come from the broken supertranslations and special sperconformal transformations.}. 

In the $\mathcal{N}$$=1$ case we can combine the $8$ zero modes to form $4$ raising operators whose application on the ground state leads to $16$ distinct states. As analyzed in \cite{Tachi} these form a single supermultiplet dubbed broken current supermultiplet. This supermultiplet contains a conserved current, and it is these that lead to an enhancement of symmetry. If additional matter is present then there are additional zero modes coming from this matter. The application of these additional modes generically charges that current also under the matter symmetry.

In the $\mathcal{N}$$=2$ case we can combine the $16$ zero modes to form $8$ raising operators whose application on the ground state leads to $256$ distinct states. Again, as analyzed in \cite{Tachi} these form a single supermultiplet whose structure is identical to that expected from the Kaluza-Klein (KK) modes of a $6d$ $\mathcal{N}$$=2$ energy-momentum supermultiplet. This is in accordance with the expectation that this theory lifts to $6d$.

For a general group $G$, the $1$ instanton is in an $SU(2)$ subgroup of $G$. Thus, to understand the behavior for general $G$ it is sufficient to decompose all representations to those of $SU(2)$ and use the preceding discussion to determine the possible states.     

Before moving to the actual calculation, a few words on the limitation of the method. First, this only tell us about conserved currents coming from the $1$ instanton. There can also be conserved currents coming from higher instantons which may lead to additional enhancement of symmetry. In some cases the need to complete a simple group necessitates the existence of additional currents with higher instanton number. Another limitation is that we only look at fermionic zero modes while the $1$ instanton spectrum also contains bosonic zero modes, and fermionic non-zero modes. With this in mind, we move to the discuss the results.

\section{The case of $USp$ groups}

In this section we discuss the case of a $USp(2N)$ gauge theory. The $1$-instanton is in an $SU(2)$ subgroup of the $USp(2N)$ gauge group, breaking it to $SU(2)\times USp(2N-2)$. Under this breaking the Adjoint of $USp(2N)$ decomposes into the Adjoint of $SU(2)$, the Adjoint of $USp(2N-2)$ and a bifundamental in the $(\bold{2},\bold{2N-2})$. As mentioned in the previous section, the $SU(2)$ Adjoint provides the fermionic zero modes to span the broken current supermultiplet. The bifundamental provides $4(N-1)$ fermionic zero modes that are in the $\bold{2N-2}$ of the unbroken $USp(2N-2)$ gauge group, and in the $\bold{2}$ of $SU_R(2)$. 

The $4(N-1)$ zero modes can be split to $\bold{2N-2}$ raising operators, $B_j$, (with charge $\frac{1}{2}$ under $U_R(1) \subset SU_R(2)$) and $\bold{2N-2}$ lowering operators, $B^{\dagger}_j$, (with charge $-\frac{1}{2}$ under $U_R(1) \subset SU_R(2)$). The full spectrum of states is generated by acting with raising operators on the ground state. However, we must also enforce $USp(2N-2)$ gauge invariance. The basic gauge invariant operator is given by $J^{ij}B_i B_j$ (where $J^{ij}$ is the invariant antisymmetric tensor of $USp(2N-2)$). Starting from the ground state, with $r=-\frac{N-1}{2}$, we can act on it with $J^{ij}B_i B_j$ operators leading to the $N$ states: 

\be
\ket{0}, J^{ij}B_i B_j\ket{0}, (J^{ij}B_i B_j)^2\ket{0}, ....., (J^{ij}B_i B_j)^{N-1}\ket{0} \label{PUSpGS}
\ee
with r-charges $-\frac{N-1}{2}, -\frac{N-1}{2}+1,.....,\frac{N-1}{2}$. 

Taking the product with the broken current supermultiplet gives a multiplet whose lowest state is a scalar with R-charge $R=\frac{N+1}{2}$. This is a BPS operator though not a conserved current. From its charges, we expect this operator to contribute $x^{N+1}$ to the index. This agrees with results found from index calculations, where the $1$ instanton contribution for a pure $USp(2N)$ gauge theory indeed behaves as: $\sim x^{N+1} + O(x^{N+2})$\cite{KKL,BGZ2}.

Finally we need also consider the residual discrete gauge symmetry. This arises as an instanton of an $SU(2)$ gauge theory does not completely break the $SU(2)$ gauge symmetry, leaving the $Z_2$ center unbroken. We must also demand gauge invariance under this symmetry. It turns out that the ground state is even for a vanishing $\theta$ parameter, and odd for $\theta=\pi$. This can be understood either directly from the path integral \cite{Tachi}, or alternatively from the exact instanton counting \cite{BGZ1}.
 
From the results for $SU(2)$ it is now straightforward to generalize to the case of $USp(2N)$. As the instanton is in an $SU(2)$ subgroup, we again conclude that the ground state is even under $Z_2$ only if $\theta=0$. We must also consider the behavior of the fermionic operators $B_i$ under this symmetry. As these arise from $SU(2)$ doublets, each one transforms under this $Z_2$ gauge symmetry. Thus, a state is even under $Z_2$ if it is made from an application of an even number of $B_i$'s when $\theta=0$, or an odd number when $\theta=\pi$. All the states in (\ref{PUSpGS}) contain an even number of $B_i$ operators and thus are kept when $\theta=0$, but are projected out when $\theta=\pi$.  

It is straightforward to generalize this result by adding $N_f$ hypers in the fundamental of $USp(2N)$. These provide $2N_f$ zero modes, which are gauge singlets, whose application on the previous states furnish it with the Dirac spinor representation of $SO(2N_f)$. Enforcing the $Z_2$ gauge invariance eliminates half of them so the end result is one state with $R=\frac{N-1}{2}$ and in a Weyl spinor representation of $SO(2N_f)$, where the chirality of the spinor is determined by the $\theta$ angle. Thus, we find no broken current supermultiplets for $USp(2N)+N_fF$. This is consistent with the expectations from instanton analysis and brane webs in the presence of orientifolds \cite{BZ1}. 

\subsection{With antisymmetric matter}

We can add matter in the antisymmetric representation of $USp(2N)$. Under the breaking of $USp(2N)\rightarrow SU(2)\times USp(2N-2)$ the antisymmetric of $USp(2N)$ decomposes to a singlet, the antisymmetric of $USp(2N-2)$, and a state in the $(\bold{2},\bold{2N-2})$. Therefore, similarly to the Adjoint, the antisymmetric provides $4N-4$ fermionic zero modes that can be combined to form $2N-2$ fermionic operators, $A_i$, in the $\bold{2N-2}$ of the unbroken gauge $USp(2N-2)$, that raise by $\frac{1}{2}$ the charge under $U_{AS}(1)\subset SU_{AS}(2)$. 

We can form a $USp(2N-2)$ invariant by $J^{ij}A_i A_j$ whose repeated application on (\ref{PUSpGS}) charges it in the $\bold{N}$ dimensional representation of $SU_{AS}(2)$. Alternatively, we can form a $USp(2N-2)$ invariant by $J^{ij}A_i B_j$, and applying this on the ground state gives a new state, as it changes the charges of both $U_{AS}(1)\subset SU_{AS}(2)$ and $U_{R}(1)\subset SU_{R}(2)$ by $\frac{1}{2}$. Additional applications of $J^{ij}A_i A_j$ and $J^{ij}B_i B_j$ charges this state in the $(\bold{N-1},\bold{N-1})$ dimensional representation of $(SU_{R}(2),SU_{AS}(2))$. Additional applications of $J^{ij}A_i B_j$ generate new states until we apply $N-1$ such operators\footnote{This hangs on the observation that the $J$ matrix always connects a specific pair of indices. For example, the combination $A_1B_2A_3B_4$ is contained in $J^{ij}A_i B_j J^{kl}A_k B_l$, but not in $J^{ij}A_i A_j J^{kl}B_k B_l$. So the states formed by applying these on the ground state are distinct. This is true until all $N-1$ possible pairs in $J^{ij}$ appear, corresponding to applying $J^{ij}A_i B_j$ $N-1$ times.}. Tensoring this with the broken current supermultiplet, we find a multitude of states in representations $(\bold{k+2},\bold{k})$ for $k=1,2..,N$, under $(SU_{R}(2),SU_{AS}(2))$. In particular this includes a state in the $(\bold{3},\bold{1})$, given by:

\be
(J^{ij}A_i B_j)^{N-1}\ket{0} \label{ASUSpGS}
\ee
which corresponds to a conserved current. 

We also need to take into account the effect of the $\theta$ angle. As all these states are made from an application of an even number of fermionic raising operators, they are kept only if $\theta=0$ while for $\theta=\pi$ they are projected out. Therefore there are no conserved currents for $\theta=\pi$.

We thus find that $USp_{0}(2N)+AS$ has a conserved current which is an $SU_{AS}(2)$ singlet suggesting an enhancement of $U_T(1)\rightarrow SU(2)$ while $USp_{\pi}(2N)+AS$ has no enhancement, at least up to the limitations of this method. This is in accordance with the identification of $USp_0(2N)+AS$ and $USp_{\pi}(2N)+AS$ as the rank $N$ $E_1$ and $\tilde{E}_1$ theory, respectively\cite{SMI}. 

It is now straightforward to generalize by adding $N_f$ fundamental flavors. As previously stated this will just charge the instanton under the spinor representation of $SO(2N_f)$. This suggests an enhancement of $U_T(1)\times SO(2N_f)$ to $E_{N_f+1}$. This again is consistent with the identification of $USp(2N)+AS+N_fF$ with the rank $N$ $E_{N_f+1}$ theory. 

Finally, we consider the case of $N_f=8$. The conserved current is now in the spinor of $SO(16)$. However, there is no finite Lie group which contains $U(1)\times SO(16)$, and whose Adjoint contains states with charges $\bold{128}^{1}, \bold{128'}^{-1}$ . On the other hand, there is an affine Lie group, $E^{(1)}_8$, that has this spectrum\footnote{We adopt the notation for affine Lie groups used in \cite{Tachi1}. The physical interpretation of this is that the group written is the $6d$ global symmetry while the superscript denote whether the reduction is done with a twist in the outer automorphism of the group ($(2)$ or $(3)$) or not ($1$).}. This suggests that this theory lifts to a $6d$ theory with $E_8$ global symmetry in agreement with the results of \cite{GMS,KKLPV}. 

We do not expect either a $5d$ or a $6d$ fixed point to exist, for general $N$, when there are two or more antisymmetrics. This is because, when $N>6$, there is a Higgs branch leading to an $SU(2)$ gauge theory with more than $1$ hyper in the Adjoint. However, for $N\leq 6$ a fixed point might exist, and in the remainder of this subsection we will deal with these cases. We start with the case of $USp(4)$ with $2$ antisymmetrics and $N_f$ flavors. This theory has a Higgs branch leading to $SU(2)+2N_fF$, so as long as $N_f<4$ a $5d$ fixed point is not ruled out.

First consider the case of $N_f=0$. We still have the previous conserved current, given by:

\be
J^{ij}A^f_i B_j\ket{0}
\ee
where we use the index $f=1,2$ for $SU(2)\subset USp_{AS}(4)$. This state is in the $\bold{2}^{-1}$ representation of $U(1)\times SU(2)\subset USp_{AS}(4)$. We can also apply $J^{ij}A^{f_1}_i A^{f_2}_j$ generating an additional conserved current, now in the $\bold{2}^{1}$ representation. These two form the $\bold{4}$ of $USp_{AS}(4)$. This conserved current is kept provided $\theta=0$. In that case, we expect an enhancement of $U_T(1)\times USp_{AS}(4)\rightarrow USp(6)$. Note that this enhancement also requires a $USp_{AS}(4)$ singlet conserved current with instanton charge $\pm 2$. 

Adding flavor will give it also charges under the spinor of $SO(2N_f)$. When $N_f=1$ the global symmetry associated with the flavor is $U_F(1)$ and the conserved current acquires the charge $\pm\frac{1}{2}$ depending on the value of the $\theta$ angle. Thus, we still expect an enhancement of $U(1)\times USp_{AS}(4)\rightarrow USp(6)$, just with the $U(1)$ being a combination of $U_T(1)$ and $U_F(1)$. 

When $N_f=2$ the global symmetry associated with the flavors is $SU(2)\times SU(2)$, and the conserved currents are in the $(\bold{2},\bold{1})$ or $(\bold{1},\bold{2})$, depending on the $\theta$ angle. The smallest global symmetry consistent with this is $USp(8)$. This also requires conserved currents with instanton number $\pm 2$ and in the $(\bold{1},\bold{3})$ of $USp_{AS}(4)\times SU(2)$. Assuming these states are indeed present, the global symmetry of this theory is $SU(2)\times USp(8)$.

When $N_f=3$ the global symmetry associated with the flavors is $SU(4)$, and the conserved currents are in the $\bold{4}$ or $\bar{\bold{4}}$, depending on the $\theta$ angle. The smallest global symmetry consistent with this is $USp(12)$. This also requires a conserved current, and it's conjugate, with instanton number $2$ and in the $(\bold{1},\bold{10})$ of $USp_{AS}(4)\times SU(4)$. Assuming these states are indeed present, the global symmetry of this theory is $USp(12)$.

Finally, we consider the case of $N_f=4$. The global symmetry associated with the flavors is now $SO(8)$, and the conserved currents are in the $\bold{8}_S$ or $\bold{8}_C$, depending on the $\theta$ angle. There is no finite Lie group, containing $U(1)\times USp(4)\times SO(8)$, whose Adjoint contains $(\bold{4},\bold{8})^{\pm 1}$. However, this is contained in the affine Lie group $A^{(2)}_{11}$ So this theory may lift to $6d$. 

There are two other theories in this family that may have a $6d$ fixed point. One is $USp(4)+3AS=SO(5)+3V$ which we deal with in the next section when we discus $SO(N)$ groups with vector matter. The second is $USp(6)+2AS$ where we find a conserved current in the $\bold{10}$ of $USp_{AS}(4)$ which is kept if $\theta=0$. This cannot span any finite Lie group rather leading to the affine $C^{(1)}_2$. 

\subsection{With symmetric matter}

In this subsection we deal with adding a hypermultiplet in the symmetric representation. For $USp$ groups, the symmetric representation is the Adjoint so this theory is the maximally supersymmetric $USp$ theory and is expected to lift to $6d$. Under the breaking $USp(2N)\rightarrow SU(2)\times USp(2N-2)$ the symmetric of $USp(2N)$ decomposes to symmetrics of $SU(2)$ and $USp(2N-2)$, and a state in the $(\bold{2},\bold{2N-2})$. 

The only difference between this case and the case with the antisymmetric is that we also have a state in the Adjoint of $SU(2)$. As mentioned in the previous section, the $SU(2)$ Adjoint hyper contributes additional zero modes generating a KK mode energy-momentum supermultiplet. We next need to take into account the effect of the bifundamental zero modes. As these are the same as in the previous subsection, we can just borrow the results so we get a tower of states in the $(\bold{k},\bold{k})$ dimensional representations of $(SU_{R}(2),SU_{S}(2))$, for $k=1,2..,N$. 

The groups $SU_{R}(2)\times SU_{S}(2)$ form a larger group $SO_R(5)$, which is the R-symmetry of the maximally supersymmetric theory, and these states form a single representation of $SO_R(5)$ which is the $N-1$ symmetric traceless representation. The final result is given by tensoring these two states. Finally, we need to consider the effect of the $\theta$ angle. If $\theta=0$ then these states are kept, but are projected out when $\theta=\pi$.

For $\theta=0$ the resulting spectrum is consistent with the expected $6d$ lift. We remind the reader that $5d$ maximally supersymmetric $USp_{0}(2N)$ theory is expected to lift to the $6d$ $(2,0)$ theory of type $D_{N+1}$ where the compactification is done with a twist in the outer automorphism of $SO(2N+2)$\cite{Tachi1}. The $(2,0)$ theory of type $D_{N+1}$ has a short multiplet for every $SO(2N+2)$ invariant polynomial. This short multiplet contains a symmetric traceless spacetime tensor that is in the $D-2$ symmetric traceless representation of $SO_R(5)$ where $D$ is the degree of the corresponding polynomial.

For $SO(2N+2)$, there are $N+1$ invariant polynomials of degrees $2,4,...,2N-2,2N$ and $N+1$. When compactified, without a twist, on a circle we get the $5d$ maximally supersymmetric $SO(2N+2)$ theory. These short multiplets get expanded into KK modes that contribute to the $5d$ theory. The constant modes on the circle are masseless and gives the corresponding supermultiplets in the $5d$ theory. The first exited state are massive with mass $\frac{1}{R} \sim \frac{1}{g^2_{5d}}$ which we expect to appear as instanton states. 

Now we want to take into account the effect of the outer automorphism twist on these results. Out of the $N+1$ polynomials, the $N$ polynomials of degrees $2,4,...,2N-2,2N$ are even while the one of degree $N+1$ is odd. Thus, with the twist, we must enforce anti-periodic boundary conditions on the supermultiplet corresponding to the degree $N+1$ invariant polynomial. As a result only those of degrees $2,4,...,2N-2,2N$ give massless states. These exactly match the degrees of the invariant polynomials of $USp(2N)$. 

The first massive state should now be the first term in the KK expansion of the degree $N+1$ invariant polynomial, and is expected to appear in the gauge theory as a $1$ instanton. It is also clear from the $6d$ reduction that, in $5d$, this state should be in the representation given by tensoring the previously mentioned KK states with a state in the $N-1$ symmetric traceless representation of $SO_R(5)$. This indeed matches the results we find from the $1$ instanton. We expect the first KK modes of the degree $2,4,...,2N-2,2N$ polynomials to contribute in the $2$ instanton.        

The $5d$ $USp_{\pi}(2N)$ maximally supersymmetric theory is expected to lift to the $6d$ $(2,0)$ theory of type $A_{2N}$ with an outer automorphism twist along the circle. In general there are $2N$ invariant polynomials of $SU(2N+1)$ having degrees $2,3...,2N,2N+1$. Under the action of the $Z_2$ outer automorphism of $SU(2N+1)$ the even degree polynomials are even while the odd degree ones are odd. Thus, we expect the first massive states to come from the degrees $3,5,...,2N-1 ,2N+1$ operators which are sensitive to the twist. Nevertheless, we find no states coming from the $1$ instanton in this case.  

\subsection{Rank $3$ antisymmetric tensor}

In this subsection we deal with other matter representations. The only other representations that seem to allow $5d$ fixed points is a rank $3$ antisymmetric tensor. Going over the possible cases we seem to find $5$ possibilities, $4$ with gauge group $USp(6)$ and one with gauge group $USp(8)$. The rank $3$ antisymmetric representation of $USp(2N)$ is pseudoreal so one can add a half-hyper. Specifically for $USp(6)$, the representation is $14$ dimensional and because of the anomaly of \cite{Wit} adding a half-hyper is inconsistent unless one also adds a half-hyper in the fundamental. The possible candidates for a $5d$ or $6d$ fixed points are $\frac{1}{2}, 1$ and $\frac{3}{2}$ rank $3$ antisymmetric hypers with fundamental hypers and a rank $3$ antisymmetric half-hyper with a rank $2$ antisymmetric and fundamental hypers. 

Before discussing each one in turn we comment about the effect of half-hypers. Under the breaking $USp(6)\rightarrow SU(2)\times USp(4)$, the $\bold{14'}$ decomposes into states in the $(\bold{1},\bold{4})$ and $(\bold{2},\bold{5})$ of $SU(2)\times USp(4)$. Together with the additional half-hyper in the fundamental, which gives fermionic zero modes in the $(\bold{2},\bold{1})$, these give $3$ fermionic raising operators whose application on the ground state charge it in the $\bold{4} \oplus \bold{4}$ of the $USp(4)$ gauge symmetry. Out of these two, only one is kept while the other is projected out, depending on the $\theta$ angle.

Now we can discuss the possible cases. First if we have just a half-hyper in the $\bold{14}'$ and the $\bold{6}$ then we essentially just need to repeat the original analysis, now with a ground state that is in the fundamental of $USp(4)$. It is not difficult to see that there are just two $USp(4)$ invariant states given by:

\be
J^{ij}B_i\ket{0}_j, J^{kl}B_k B_l J^{ij}B_i\ket{0}_j
\ee
These form a doublet of $SU_R(2)$ so there are no conserved currents. 

The generalization by adding $N_f$ fundamental hypers is immediate. Like in previous cases, the additional zero modes charge this state with the Dirac spinor representation of $SO(2N_f)\subset SO(2N_f+1)$. Enforcing $Z_2$ gauge invariance reduce it to a Weyl spinor of $SO(2N_f)$. However, we now recall that there is an additional ground state that is not invariant under $Z_2$. When flavors are added, this ground state also contributes a Weyl spinor of $SO(2N_f)$, but with opposite chirality. These combine to form an $SO(2N_f+1)$ spinor. So the end result is we find an $SU_R(2)$ doublet that is in the spinor representation of $SO(2N_f+1)$. This does not give a conserved current. 

If we add a rank $2$ antisymmetric then we need to repeat the analysis of section $3.1$ with a ground state in the fundamental of $USp(4)$. It is not difficult to see that there are just two conserved currents, given by:

\be
(J^{kl}A_k B_l)^2 J^{ij}A_i\ket{0}_j, J^{kl}A_k B_l J^{ij}B_i\ket{0}_j,
\ee
which form a doublet of $SU_{AS}(2)$. The minimal enhancement consistent with this is $U_T(1)\times SU_{AS}(2)\rightarrow SU(3)$. We can further generalize by adding flavors in the fundamental of $USp(4)$. The Higgs branch analysis suggests that we can add up to $2$ before the theory is expected to lift to $6d$. The addition of the flavors will still give the same conserved current, but now it will also be in the spinor of $SO(2N_f+1)$.

Thus, if $N_f=1$ we get a conserved current in the $(\bold{2},\bold{2})$ of $SU_{AS}(2)\times SO_F(3)$. The minimal enhancement consistent with this is $U_T(1)\times SU_{AS}(2) \times SO_F(3)\rightarrow SU(4)$. If $N_f=2$ we get a conserved current in the $(\bold{2},\bold{4})$ of $SU_{AS}(2)\times SO_F(5)$. The minimal enhancement consistent with this is $U_T(1)\times SU_{AS}(2) \times SO_F(5)\rightarrow USp(8)$ which also requires conserved currents in the $\bold{3}$ of $SU_{AS}(2)$ with instanton number $\pm 2$. Finally, if $N_f=3$ we get a conserved current in the $(\bold{2},\bold{8})$ of $SU_{AS}(2)\times SO_F(7)$. Such a spectrum cannot fit in a finite Lie group, but can fit in the affine Lie group $E^{(2)}_6$.

Next we turn to the case of one full hyper in the $\bold{14}'$ of $USp(6)$. We find two conserved currents with charges $\pm\frac{3}{2}$ under $U_{TAS}(1)$. Depending on the $\theta$ angle, one of them is projected out while the other is kept. Thus, this suggests an enhancement of $U(1)\rightarrow SU(2)$. When flavors are added we still get the two conserved currents, but now both are in the Dirac spinor representation of $SO(2N_f)$. Half of these are projected out leaving two $SO(2N_f)$ spinors of opposite chirality and with charges $\pm\frac{3}{2}$ under $U_{TAS}(1)$.

 A Higgs branch analysis suggests that we can have at most $4$ flavors and still have a $5d$ fixed point. For $N_f=1$ we get two conserved currents with opposite charges under $SO_F(2)$ and $U_{TAS}(1)$. The minimal enhancement consistent with this is $U(1)^2\rightarrow SU(2)^2$. For $N_f=2$ we get two conserved currents of opposite charges under $U_{TAS}(1)$, one in the $\bold{2}$ of one $SU_F(2)$ and the other in the $\bold{2}$ of the other. The minimal enhancement consistent with this is $U(1)^2\times SU(2)^2\rightarrow SU(3)^2$. For $N_f=3$ We get two conserved currents in the $\bold{4}$ and $\bar{\bold{4}}$ of $SU_F(4)$. The minimal enhancement consistent with this is $U(1)^2\times SU_F(4) \rightarrow SU(6)$ which also requires two flavor singlet conserved currents with instanton number $\pm 2$.

For $N_f=4$ we get two conserved currents, one in the $\bold{8}_S$ and the other in the $\bold{8}_C$ of $SO_F(8)$. This suggests an enhancement of $U(1)^2\times SO(8)\rightarrow E_6$ assuming we also get two conserved currents with instanton number $\pm 2$ and in the $\bold{8}^0_V$ of $SO(8)^{U_{TAS}(1)}$. Finally, we consider the case of $N_f=5$. In this case we get two conserved currents in the $\bold{16}$ and $\bar{\bold{16}}$ of $SO_F(10)$. This cannot fit in a finite Lie group, but can fit in the affine Lie group $E^{(1)}_6$.

Next, we consider the case of $\frac{3}{2}$ hypers in the rank $3$ antisymmetric. We do not find any conserved current in this case.

The last case we consider is $USp(8)$ with a rank $3$ antisymmetric. The Higgs branch analysis suggests we can only put a single half-hyper. This theory does not suffer from the anomaly of \cite{Wit} (see for example \cite{BT}). The effect of the half-hyper is to furnish the ground state in the $\bold{64}$ of the unbroken $USp(6)$ gauge symmetry. We find no conserved current.    
 
\section{The case of $SO$ groups}

In this section we discus the case of $SO(N)$ gauge theory. The $1$-instanton is in an $SU(2)$ subgroup of the $SO(N)$ gauge group breaking $SO(N)$ to $SU(2)\times SU(2)\times SO(N-4)$. Under this breaking the Adjoint of $SO(N)$ decomposes into the Adjoints of both $SU(2)$'s, the Adjoint of $SO(N-4)$ and a state in the $(\bold{2},\bold{2},\bold{N-4})$. Like in the previous case, the Adjoint builds a broken current supermultiplet while the remaining states provide additional fermionic zero modes. In the case at hand these are $4(N-4)$ states in the $(\bold{2},\bold{N-4})$ of the unbroken $SU(2)\times SO(N-4)$ gauge group, and the $\bold{2}$ of $SU_R(2)$.

These naturally form $2(N-4)$ raising operators $B_{a j}$ (where $a=1,2 \in SU(2), i=1,...,N-4\in SO(N-4)$) from which we can form a gauge invariant by: $\epsilon^{a b} \delta^{i j} B_{a i} B_{b j}$. 
Starting from the ground state, with $r=-\frac{N-4}{2}$, and acting with these operators, leads to the $N-3$ states:

\be
\ket{0}, \epsilon^{a b} \delta^{i j} B_{a i} B_{b j}\ket{0}, (\epsilon^{a b} \delta^{i j} B_{a i} B_{b j})^2\ket{0}, ....., (\epsilon^{a b} \delta^{i j} B_{a i} B_{b j})^{N-4}\ket{0} \label{PSOpGS}
\ee 
whose r-charges are: $-\frac{N-4}{2}, -\frac{N-4}{2}+1,.....,\frac{N-4}{2}$. Thus, we find a single ground state with R-charge $R=\frac{N-4}{2}$. Tensoring with the broken current supermultiplet we do not get a conserved current, rather a BPS operator which we expect to contribute $x^{N-2}$ to the index. Indeed, this matches the contribution one finds from the explicit index calculation. 

Next, we generalize by adding $N_f$ hypers in the vector representation. These add $4N_f$ fermionic zero modes in the $(\bold{2},\bold{1})$ of the unbroken $SU(2)\times SO(N-4)$ gauge group, and the $\bold{2N_f}$ of the $USp(2N_f)$ flavor symmetry. From these we can form $2N_f$ raising operators $C_{a}$ in the $(\bold{2},\bold{1})$ of the unbroken gauge symmetry and $\bold{N_f}^1$ under the $U(1)\times SU(N_f) \subset USp(2N_f)$ flavor symmetry.

 From these we can form the gauge invariant $\epsilon^{a b} C_{a} C_{b}$. Applying this on (\ref{PSOpGS}) yields a new state which is in the rank $2$ symmetric representation of $SU(N_f) \subset USp(2N_f)$. Repeated application of $\epsilon^{a b} C_{a} C_{b}$ leads to additional states. Their $SU(N_f) \subset USp(2N_f)$ representation should, on the one hand, be given by a symmetric product of the rank $2$ symmetric representation of $SU(N_f)$, yet on the other hand, since the underlying operators are fermionic, it cannot be a completely symmetric product. The easiest way to determine the representation is to use the fact that the state given by applying $(\epsilon^{a b} C_{a} C_{b})^l$ should be the conjugate of the one given by applying $(\epsilon^{a b} C_{a} C_{b})^{N_f-l}$. We end up with the $N_f+1$ states:

 \be
\ket{B}, \epsilon^{a b} C_{a} C_{b}\ket{B}, (\epsilon^{a b} C_{a} C_{b})^2\ket{B}, ....., (\epsilon^{a b} C_{a} C_{b})^{N_f}\ket{B} \label{FlSOpGS}
\ee  
where we have collectively denoted the states in (\ref{PSOpGS}) as $\ket{B}$. Their charges under $U(1)\subset USp(2N_f)$ are $-N_f,-N_f+2,...,N_f$ while their $SU(N_f) \subset USp(2N_f)$ representation is given by $1$,{\tiny\yng(2)},{\tiny\yng(2,2)},{\tiny\yng(2,2,2)},....., where the last Young diagram has $N_f$ rows.

We next need to combine them into $USp(2N_f)$ representations. We claim that these states build the rank $N_f$ irreducible antisymmetric representation of $USp(2N_f)$. Recall that under the $U(1)\times SU(N_f)$ subgroup of $USp(2N_f)$, the $\bold{2N_f}$ dimensional representation decomposes as $\bold{N_f}^1 + \bar{\bold{N_f}}^{-1}$, and that all $USp(2N_f)$ representations can be build from products of the fundamental representation. In order to get a state with lowest $U(1)$ charge $-N_f$ we must multiply $N_f$ fundamentals. Furthermore, for that state to be an $SU(N_f)$ singlet the product must be completely antisymmetric. Also one can count the number of states, using Mathematica for example, and show that their number exactly match the dimension of the rank $N_f$ irreducible antisymmetric representation of $USp(2N_f)$.

In addition we may be able to form an invariant from both $C_{a}$ and $B_{a j}$. Since only the $B$'s are charged under $SO(N-4)$ they must independently combine to form an invariant under that group. This can be done in two ways. First, we can contract two $B$'s using $\delta^{i j}$, but, as this is a symmetric product and the $B$'s are fermionic, the gauge $SU(2)$ indices must be contracted antisymmetrically forming a gauge invariant. This leads to the previously discussed operator. The second option is to contract $N-4$ indices with the epsilon tensor. As this is an antisymmetric product, the gauge $SU(2)$ indices must now be contracted symmetrically forming the $N-3$ dimensional representation of $SU(2)$. In order to get a different state we need to contract with something made from the $C$'s. To get the $N-3$ dimensional representation of $SU(2)$, so that it can form an $SU(2)$ gauge invariant, we must contract $N-4$ $C$'s symmetrically. Yet, as the $C$'s are fermionic they must be antisymmetrized in the flavor index. 

Thus, we conclude that when $N_f<N-4$ there are not enough $C$'s to form this invariant and the $1$-instanton comprises one state with R-charge $\frac{N-2}{2}$ and in the rank $N_f$ irreducible antisymmetric representation of $USp(2N_f)$. 

When $N_f=N-4$, in addition to the previously mentioned state, there is an invariant made from $N-4$ $B$'s and $N-4$ $C$'s given by:

\be
\epsilon^{a_1 b_1}....\epsilon^{a_{N-4} b_{N-4}} \epsilon^{i_1....i_{N-4}} \epsilon_{f_1....f_{N-4}}  B_{a_1 i_1}....B_{a_{N-4} i_{N-4}} C^{f_1}_{b_1}....C^{N-4}_{b_{N-4}}\ket{0},
\ee
where we used $f=1,...N_f$ for the $SU(N_f) \subset USp(2N_f)$ index. This is a singlet under $USp(2N_f)$ with R-charge zero and so leads to a conserved current multiplet when tensored with the basic broken current multiplet. This should lead to an enhancement of $U_T(1)\rightarrow SU(2)$. 

When $N_f=N-3$ this state acquires charges in the fundamental of $USp(2N_f)$. The minimal symmetry containing $U(1)\times USp(2N_f)$ with additional states in the $\bold{2N_f}^{\pm 1}$ is $USp(2N_f+2)$. This requires also two $USp(2N_f)$ singlets with instanton charge $\pm 2$. The enhancement spectrum seen here is consistent with the results from the explicit index analysis and brane webs\cite{BZ1}.

Finally when $N_f=N-2$ this state acquires charges in the rank $2$ irreducible antisymmetric representation of $USp(2N_f)$. There is no finite Lie group with this content, but there is an affine Lie group, $A^{(2)}_{2N_f-1}$, with this spectrum. This is in accordance with the Higgs branch analysis suggesting that this theory doesn't possess a $5d$ fixed point, though it may possess a $6d$ one. 

\subsection{With antisymmetric matter}

In this section we consider an $SO(N)$ gauge theory with a single hyper in the antisymmetric representation. Since for $SO$ groups the antisymmetric is the Adjoint representation, this is the maximally supersymmetric case. The zero modes contributed by the antisymmetric hyper are $8$ zero modes coming from an $SU(2)$ Adjoint, and $4(N-4)$ zero modes coming from doublets of $SU(2)$ that are also in the $(\bold{2},\bold{N-4})$ of the unbroken $SU(2)\times SO(N-4)$ gauge group. The effect of the $8$ zero modes coming from the $SU(2)$ Adjoint hypermultiplet are just to build the previously mentioned KK modes.

The other $4(N-4)$ zero modes can be combined to form $2(N-4)$ raising operators, $A_{a i}$, that are in the $(\bold{2},\bold{N-4})$ of the unbroken $SU(2)\times SO(N-4)$ gauge group. These can be combined with the previous $B_{a i}$ operators to form the operator $B^{\alpha}_{a i}$ (where $\alpha=1,2$ and $B^{1}_{a i}=B_{a i}$, $B^{2}_{a i}=A_{a i}$). This is more than just a notational convenience as $SU_R(2) \times SU_{AS}(2)$ is enhanced to $SO_R(5)$ of the maximally supersymmetric theory. In addition to its gauge charges, The operator $B^{\alpha}_{a i}$ is charged in the $\bold{2}^{1}$ of $U(1)\times SU(2)\subset SO_R(5)$. Note that this $SU(2)$ is neither $SU_R(2)$ nor $SU_{AS}(2)$, rather a different $SU(2)$ subgroup of $SO_R(5)$. Its defining property is that the $\bold{5}$ of $SO_R(5)$ decomposes to the $\bold{1}^2 + \bold{1}^{-2} + \bold{3}^{0}$ of $U(1)\times SU(2)\subset SO_R(5)$. 

 We next need to apply these zero modes on the ground states, enforcing $SU(2)\times SO(N-4)$ invariance, and compose the results into $SO_R(5)$ representations. As previously stated the basic gauge invariant one can make is $\delta^{i j} \epsilon^{a b} B^{\alpha}_{a i} B^{\beta}_{b j}$. Due to the symmetry properties of the involved operators, this operator is in the $\bold{3}^2$ of $U(1)\times SU(2)\subset SO_R(5)$. By repeatedly applying it on the ground states we generate the states:

\be
\ket{0},\delta^{i j} \epsilon^{a b} B^{\alpha}_{a i} B^{\beta}_{b j}\ket{0}, (\delta^{i j} \epsilon^{a b} B^{\alpha}_{a i} B^{\beta}_{b j})^2\ket{0}.... (\delta^{i j} \epsilon^{a b} B^{\alpha}_{a i} B^{\beta}_{b j})^{2(N-4)}\ket{0}. \label{SOwASM}
\ee

These states are in the $\bold{1}^{-2(N-4)},$ $\bold{3}^{-2(N-5)}, \otimes^{2}_{sym}\bold{3}^{-2(N-6)}$ $,..., \otimes^{N-5}_{sym}\bold{3}^{-2},$ $\otimes^{N-4}_{sym}\bold{3}^{0},$ $\otimes^{N-5}_{sym}\bold{3}^{2},$ $ ....,\bold{3}^{2(N-5)}, \bold{1}^{2(N-4)}$ representation of $U(1)\times SU(2)\subset SO_R(5)$ (where $\otimes^{l}_{sym}$ stands for the $l$ symmetric product of the representation with itself). This in fact forms a single state in the rank $N-4$ symmetric traceless representation of $SO_R(5)$. This can be seen by starting with the decomposition $\bold{5}=\bold{1}^2 + \bold{1}^{-2} + \bold{3}^{0}$, and doing the symmetric multiplication. After removing the trace, one can see that we get the spectrum of (\ref{SOwASM}). 

Alternatively, we can contract the $SU(2)\subset SO_R(5)$ and $SO(N-4)$ indices antisymmetrically. This does not give an $SO(N-4)$ invariant, but we can form one by a symmetric product of two such operators. In term of the $B$ operators, it is given by $\delta^{i k} \delta^{j l}\epsilon^{a b} \epsilon_{\alpha \beta} B^{\alpha}_{a i} B^{\beta}_{b j} \epsilon^{c d} \epsilon_{\gamma \delta} B^{\gamma}_{c k} B^{\delta}_{d l}$.

 Acting with this operator on the ground state generates a new state, and repeated application of the operator $\delta^{i j} \epsilon^{a b} B^{\alpha}_{a i} B^{\beta}_{b j}$ on it generates a single state in the rank $N-6$ symmetric traceless representation of $SO_R(5)$. It is now clear that operating on the ground state with these two types of gauge invariants generates a series of states in the rank $N-4, N-6, N-8...$ symmetric traceless representations of $SO_R(5)$. For $N$ odd, these are the only gauge invariant states, but for $N$ even, there is one more. 

We can form an $SO(N-4)$ invariant by contracting $N-4$ $B$ operators with an epsilon tensor, and if $N$ is even, we can contract their $SU(2)$ indices among themselves so as to form an invariant. This comprises a new state, and again by repeated application of $\delta^{i j} \epsilon^{a b} B^{\alpha}_{a i} B^{\beta}_{b j}$ we get a single state in the rank $\frac{N-4}{2}$ symmetric traceless representation of $SO_R(5)$.

So, to recapitulate, for $N$ odd we find a list of states in the rank $N-4, N-6, N-8...,3,1$ symmetric traceless representations of $SO_R(5)$. However, for $N$ even, we get a list of states in the rank $N-4, N-6, N-8...,2,0$ and $\frac{N-4}{2}$ symmetric traceless representations of $SO_R(5)$. 

Next, we compare this against the expectations from the reduction of the corresponding $(2,0)$ theory. As stated in the previous section, in the $N$ even case the theory is expected to lift to the $6d$ $(2,0)$ of type $D_{\frac{N}{2}}$, and the short multiplets of this theory are expected to give KK modes that are precisely in the rank $N-4, N-6, N-8...,2,0$ and $\frac{N-4}{2}$ symmetric traceless representations of $SO_R(5)$. 

For the $N$ odd case, the theory is expected to lift to the $6d$ $(2,0)$ of type $A_{N-1}$ with a $Z_2$ twist along the circle. Thus, the operators corresponding to the odd degree polynomials obey anti-periodic boundary conditions on the circle. The even ones contribute at the massless level matching the degree $2,4,..,N-1$ invariant polynomials of $SO(N)$ for odd $N$. The first massive states correspond to the lowest KK mode of operators corresponding to the odd degree polynomials. This exactly matches our results for the $1$ instanton contribution.  

\subsection{With spinor matter}

Next we consider the generalization by the addition of spinor matter. This is especially interesting as conventional instanton counting is unavailable in this case. Generically, under the breaking $SO(N)\rightarrow SO(4)\times SO(N-4)$, the spinor of $SO(N)$ decomposes into two bispinors. So we are to determine whether, by applying all possible fermionic zero modes and limiting to gauge invariant states, there exists an R-charge singlet, and with what charges under the flavor symmetry. As this can be quite involved in general, we have resorted to numerical methods in many cases. 

The numerical strategy we use borrows significantly from index calculations. First, we pack all fermionic zero modes in a one particle index, defined by:

\be
I_{OPI} = -\sum_i x_i \chi_i ,
\ee
where the sum goes over all types of fermionic zero modes, those coming from the Adjoint, vector matter and spinor matter. Here, $x_i$ stands for the fugacity of the global $U(1)$ raised by the corresponding fermionic zero mode. Finally, we collectively denote by $\chi_i$ the character of that type of zero modes under the non-abelian flavor and gauge symmetries. The minus sign is inserted as these are fermionic operators. Next, the one particle index is inserted into a plethystic exponent\footnote{The plethystic exponent is defined as $PE[f(\cdot)] = exp\left[\sum^{\infty}_{n=1} \frac{1}{n} f(\cdot^n)\right]$ where the dot represents all the variables in $f$.}, which is expanded in a power series in all $x_i$'s. This generates all possible products of these zero modes taking into account their fermionic nature. Next, we need to act with these operators on the ground state which in practice amounts to multiplying by the charges carried by the ground state.

All that remains is to enforce gauge invariance, which is done by integrating over the gauge group with the appropriate Haar measure. From the final result we can identify whether there are $SU_R(2)$ singlets, and in what representation of the global symmetry.

As the properties of spinors vary with $N$ we concentrate on specific cases. Note, that the cases of $N=3,5$ were already covered in the previous cases. The case of $N=4$ is not a simple group and so won't be discussed here. Finally, the case of $SO(6)=SU(4)$ with only spinor matter was covered in \cite{Tachi}. While, to our knowledge, the case with both vector and spinor matter was not addressed, we won't discuss it here. Thus, the first case we discuss is $SO(7)$. Also, a Higgs branch analysis suggests that for $N>14$ a $5d$ fixed point is not possible, when spinor matter is present, so we won't consider these cases.

\subsubsection{SO(7)}

In this case the fermionic zero modes, provided by the spinors, are in the $(\bold{1},\bold{2})$ of the $SU(2)^2$ unbroken gauge group. Next we state our results for the conserved currents. When $N_f=4$, the maximal number of spinors is $N_s=1$ (if $N_s=2$ the theory as a Higgs branch where it reduces to $SU(2)+8F$). We recall that for $N_s=0$ there is a conserved current in the $\bold{8}$ of $USp_F(8)$. The addition of spinors also gives it charges under the global symmetry associated with the spinors. Particularly, for $N_s=1$, it acquires the representation $\bold{2}$ of $SU_S(2)$. We find no additional conserved currents, from the 1-instanton, besides this. The minimal global symmetry consistent with this is $USp(12)$. This also requires conserved currents with instanton number $\pm 2$ and in the $(\bold{1},\bold{3})$ of $USp_F(8)\times SU_{S}(2)$. Assuming these states are indeed present, the global symmetry of this theory is $USp(12)$. 

When $N_s=2$, we get a conserved current in the $(\bold{5},\bold{8})$ of $USp_S(4)\times USp_F(8)$. This cannot fit in a finite Lie group, but can form an affine one, $A^{(2)}_{12}$.

For $N_f=3$, the maximal number of spinors is $N_s=3$ (if $N_s=4$ the theory has a Higgs branch where it reduces to two copies of $SU(2)+8F$). We recall that for $N_s=0$ there is a conserved current which is a $USp_F(6)$ singlet. The addition of spinors gives it charges under the spinor global symmetry. For $N_s=1$ this state acquires charges under the $\bold{2}$ of $SU_S(2)$. The minimal global symmetry consistent with this is $SU(3)$. For $N_s=2$, the conserved current is now in the $\bold{5}$ of $USp_S(4)$. The minimal global symmetry consistent with this is $SO(7)$. When $N_s=3$, the conserved current is now in the $\bold{14}'$ of $USp_S(6)$. The minimal global symmetry consistent with this is $F_4$, which also requires two extra conserved currents with instanton number $\pm 2$.  

Finally, for $N_s=4$ the conserved current is now in the $\bold{42}$ of $USp_S(8)$. There is no finite Lie group that can accommodate this structure, but there is an affine group, $E^{(2)}_6$, that can. Also it turns out that there is an additional conserved current in the $(\bold{1},\bold{14})$ of $USp_S(8)\times USp_F(6)$. Again this appears to suggest an enhancement to an affine group $A^{(2)}_5$. Thus, in this case, both the global symmetries appear to be affinized which is consistent with this theory lifting to $6d$.

For $N_f=2$ we find no conserved currents unless $N_s\geq 4$. When $N_s= 4$, this current is in the $(\bold{1},\bold{4})$ of $USp_S(8)\times USp_F(4)$. This appears to suggest an enhancement of $U_T(1)\times USp_F(4)$ to $USp(6)$. This also requires two states with instanton number $\pm 2$. This is the maximal number of spinors allowed in this case. For $N_s=5$, there is a Higgs branch leading to $SU(2)+8F$ so one might expect this theory to lift to $6d$. Indeed, in this case, we find the conserved current to be in the $(\bold{10},\bold{4})$ of $USp_S(10)\times USp_F(4)$ which is consistent with an affine group, in this case $C^{(1)}_7$. 

For $N_f=1$ we find no conserved currents unless $N_s\geq 4$. When $N_s= 4$, this current is a singlet under the flavor symmetry. This should lead to an enhancement of $U_T(1)\rightarrow SU(2)$. For $N_s=5$ it acquires charges in the fundamental of $USp_S(10)$ suggesting an enhancement to $USp(12)$ (again this also requires states with instanton number $\pm 2$). Finally, in the case of $N_s=6$, the conserved current is in the rank $2$ irreducible antisymmetric of $USp_S(12)$. This cannot be accommodated in a finite Lie group, but rather in the affine $A^{(2)}_{11}$. This is consistent with a Higgs branch analysis, as for this case the theory can be reduced to $SU(2)+8F$. Strangely, $SU_F(2)$ does not appear to be affinized at this level.

Finally we can consider the case with no vectors. The maximal number of spinors is $6$, as for $7$ spinors there is a Higgs branch where the theory breaks to $SU(2)+8F$. To form an R-charge singlet we must act on the ground state, whose r-charge is $-\frac{3}{2}$, with three $B$ operators. However, it is not possible to form an $SU(2)$ gauge invariant from three doublets, so one cannot form a state that is both a gauge and R-charge singlet. Therefore, there are no conserved currents in this case, for any number of spinors. 

The important element here is the need to apply an odd number of $B$ operators in order to get an R-charge singlet. This is true for any $SO(M)$ with spinor matter only, for $M$ odd, so in all these cases there is no symmetry enhancement coming from the $1$-instanton.

We summarize our results for the conserved currents, and the expected symmetry enhancement, for theories with a potential $5d$ fixed point, in table \ref{summary1}.

\begin{table}[h!]
\begin{center}
\begin{tabular}{|c|c|c|c|}
  \hline 
   $N_f$ & $N_s$ & Currents & Minimal enhanced symmetry \\
 \hline
  $1$ & $4$ & $(\bold{1},\bold{1})$  & $ SU(2)\times SU_F(2)\times USp_S(8)$ \\
 \hline
$1$ & $5$ & $(\bold{1},\bold{10})$  & $SU_F(2)\times USp(12)^{*}$  \\
 \hline
$2$ & $4$ & $(\bold{4},\bold{1})$  & $USp_S(8)\times USp(6)^{*}$ \\
 \hline
$3$ & $1$ & $(\bold{1},\bold{2})$ & $USp_F(6)\times SU(3)$ \\
 \hline
$3$ & $2$ & $(\bold{1},\bold{5})$ & $USp_F(6)\times SO(7)$ \\
 \hline
$3$ & $3$ & $(\bold{1},\bold{14}')$ & $USp_F(6)\times F_4 ^{*}$  \\
 \hline
$4$ & $1$ & $(\bold{8},\bold{2})$ & $USp(12)^{**}$ \\
 \hline
\end{tabular}
 \end{center}
\caption{The enhancement of symmetry for the $5d$ theory $SO(7)+N_f\bold{7}+N_s\bold{8}$. The classical global symmetry is $U_T(1)\times USp(2N_f)\times USp(2N_s)$. Written are the found conserved currents with their representation under the $USp(2N_f)\times USp(2N_s)$ global symmetry, and minimal enhanced symmetry consistent with these currents. Only cases where conserved currents were found, and where a $5d$ fixed point is not ruled out, are shown. $*$ This enhancement also requires two conserved currents that are flavor singlets with instanton number $\pm 2$. $**$ This enhancement also requires two conserved currents that are in the $(\bold{1},\bold{3})$ with instanton number $\pm 2$.} 
\label{summary1}
\end{table}

\subsubsection{SO(8)}

Next we move to the case of $SO(8)$. There are two different, self-conjugated, spinor representations. Under the unbroken $SU(2)^3$ gauge group, the fermionic zero modes, provided by the spinors, are in the $(\bold{1},\bold{2},\bold{1})$ or $(\bold{1},\bold{1},\bold{2})$ depending on the chirality\footnote{In this notation the fermionic zero modes provided by the vector are in the $(\bold{2},\bold{1},\bold{1})$, and triality of $SO(8)$ is correctly implemented by permutating the three $SU(2)$ groups.}. Next we state our results for the conserved currents. To save space, henceforward we shall state only those cases where a conserved current was found. We also present only cases unrelated by triality, in this case, or, in other cases, the duality exchanging the two spinor representations. 

For $N_f=5$ we can have at most a single spinor of any chirality as for $N_s=2$ or $N_s=N_c=1$ there is a Higgs branch leading to $SU(2)+8F$. Recall that when $N_s=N_c=0$ there is a single conserved current in the $\bold{10}$ of $USp_F(10)$. The addition of spinor matter charge it also under the spinor flavor symmetry. Thus, for $N_s=1$ it acquires the $\bold{2}$ of $SU_S(2)$ so we find a single conserved current in the $(\bold{10},\bold{2})$ of $USp_F(10)\times SU_S(2)$. This should lead to an enhancement of $U_T(1)\times SU_S(2)\times USp_F(10)\rightarrow USp(14)$. This also requires conserved currents with instanton number $\pm 2$ and charges $(\bold{1},\bold{3})$.

The cases of $N_s=2$ and $N_s=N_c=1$ are also interesting as they may possess a $6d$ fixed point. For $N_s=2$, we find a conserved current in the $(\bold{10},\bold{5})$ of $USp_F(10)\times USp_S(4)$. This does not fit in a finite Lie group, but rather in the affine $A^{(2)}_{14}$. For $N_s=N_c=1$, we find a conserved current in the $(\bold{10},\bold{2},\bold{2})$ of $USp_F(10)\times SU_S(2)\times SU_C(2)$. This again does not fit in a finite Lie group, but rather in the affine $A^{(2)}_{13}$.

When $N_f=4$, the maximal possible number of spinors, in any combination of chiralities, is three, as the case of four spinors has a Higgs branch leading to two copies of $SU(2)+8F$. Recall that when $N_s=N_c=0$ there is a conserved current which is a $USp_F(8)$ singlet. The addition of spinor matter charges this current under the spinor flavor symmetry. For the $N_s=1, N_c=0$ case, it acquires the $\bold{2}$ of $SU_S(2)$. Thus, we find a conserved current in the $(\bold{1},\bold{2})$ of $USp_F(8)\times SU_S(2)$. This should lead to an enhancement of $U_T(1)\times SU_S(2)\rightarrow SU(3)$. For $N_s=2, N_c=0$,  it acquires the $\bold{5}$ of $USp_S(4)$ and we find a conserved current in the $(\bold{1},\bold{5})$ of $USp_F(8)\times USp_S(4)$. This should lead to an enhancement of $U_T(1)\times USp_S(4)\rightarrow SO(7)$.

Alternatively, for $N_s=N_c=1$, we find a conserved current in the $(\bold{1},\bold{2},\bold{2})$ of $USp_F(8)\times SU_S(2)\times SU_C(2)$. This should lead to an enhancement of $U_T(1)\times SU_S(2)\times SU_C(2)\rightarrow SU(4)$. For $N_s=3, N_c=0$, we find a conserved current in the $(\bold{1},\bold{14}')$ of $USp_F(8)\times USp_S(6)$. This should lead to an enhancement of $U_T(1)\times USp_S(6)\rightarrow F_4$, which also requires conserved currents with instanton number $\pm 2$ that are flavor symmetry singlets. The last remaining case is $N_s=2, N_c=1$ for which the spinor fermionic zero modes charge the current in the $(\bold{5},\bold{2})$ of $USp_S(4)\times SU_C(2)$. So we find a conserved current in the $(\bold{1},\bold{5},\bold{2})$ of $USp_F(8)\times USp_S(4)\times SU_C(2)$. This should lead to an enhancement of $U_T(1)\times USp_S(4)\times SU_C(2)\rightarrow SO(9)$, which also requires conserved currents with instanton number $\pm 2$ that are flavor symmetry singlets.

The cases with four spinors are also interesting as they may have a $6d$ fixed point. There are three relevant cases. For $N_s=4, N_c=0$, we find two conserved currents, one in the $(\bold{42},\bold{1})$, and the other in the $(\bold{1},\bold{42})$ of $USp_F(8)\times USp_S(8)$. These appear to affinize both $USp(8)$'s to $E^{(2)}_{6}$. For $N_s=3, N_c=1$, we find a conserved current in the $(\bold{1},\bold{14}',\bold{2})$ of $USp_F(8)\times USp_S(6)\times SU_C(2)$. This should affinize $USp_S(6)\times SU_C(2)$ to $F^{(1)}_4$. Incidentally, $USp_F(8)$ is not affinized, at least not at this level. Finally for  $N_s=2, N_c=2$, we find two conserved currents, one in the $(\bold{1},\bold{5},\bold{5})$, and the other in the $(\bold{27},\bold{1},\bold{1})$ of $USp_F(8)\times USp_S(4)\times USp_C(4)$. These appear to affinize $USp_F(8)$ to $A^{(2)}_{7}$ and $USp_S(4)\times USp_C(4)$ to $D^{(2)}_5$. 

There are just two other cases, not related by triality to the previous cases, where we find a conserved current. The first is $N_f=N_s=N_c=2$ where we find a single conserved current which is a flavor singlet. This should lead to an enhancement of $U_T(1)\rightarrow SU(2)$. The second being $N_f=3, N_s=N_c=2$ where we find a conserved current in the $(\bold{6},\bold{1},\bold{1})$ of $USp_F(6)\times USp_S(4)\times USp_C(4)$. This should lead to an enhancement of $U_T(1)\times USp_F(6)\rightarrow USp(8)$, which also requires conserved currents with instanton number $\pm 2$ that are flavor symmetry singlets.

The case of $N_f=N_s=3, N_c=2$ is also interesting as in that case there is a Higgs branch leading to $SU(2)+8F$ so while we do not expect a $5d$ fixed point, a $6d$ one is possible. Indeed, we find a conserved current in the $(\bold{6},\bold{6},\bold{1})$ of $USp_F(6)\times USp_S(6)\times USp_C(4)$. This should affinize $USp_F(6)\times USp_S(6)$ to $C^{(1)}_6$. Incidentally, $USp_C(4)$ is not affinized, at least not at this level.

We summarize our results for the conserved currents, and the expected symmetry enhancement, for theories with a potential $5d$ fixed point, in table \ref{summary2}.

\begin{table}[h!]
\begin{center}
\begin{tabular}{|c|c|c|c|c|}
  \hline 
   $N_f$ & $N_s$ & $N_c$ & Currents & Minimal enhanced symmetry \\
 \hline
  $2$ & $2$ & $2$ & $(\bold{1},\bold{1},\bold{1})$  & $SU(2)\times USp_F(4) \times USp_S(4) \times USp_C(4)$ \\
 \hline
$3$ & $2$ & $2$ & $(\bold{6},\bold{1},\bold{1})$  & $USp_S(4) \times USp_C(4) \times USp(8)^{*}$ \\
 \hline
$4$ & $1$ & $0$ & $(\bold{1},\bold{2})$ & $USp_F(8)\times SU(3)$  \\
 \hline
$4$ & $1$ & $1$ & $(\bold{1},\bold{2},\bold{2})$  & $USp_F(8)\times SU(4)$ \\
 \hline
$4$ & $2$ & $0$ & $(\bold{1},\bold{5})$ & $USp_F(8)\times SO(7)$ \\
 \hline
$4$ & $2$ & $1$ & $(\bold{1},\bold{5},\bold{2})$ & $USp_F(8)\times SO(9)^{*}$ \\
 \hline
$4$ & $3$ & $0$ & $(\bold{1},\bold{14}')$ & $USp_F(8)\times F_4 ^{*}$  \\
 \hline
$5$ & $1$ & $0$ & $(\bold{10},\bold{2})$ & $USp(14)^{**}$ \\
 \hline
\end{tabular}
 \end{center}
\caption{The enhancement of symmetry for the $5d$ theory $SO(8)+N_f\bold{8}_V+N_s\bold{8}_S+N_c\bold{8}_C$. The classical global symmetry is $U_T(1)\times USp(2N_f)\times USp(2N_s)\times USp(2N_c)$. Written are the found conserved currents with their representation under the $USp(2N_f)\times USp(2N_s)\times USp(2N_c)$ global symmetry, and minimal enhanced symmetry consistent with these currents. Only cases where conserved currents were found, and where a $5d$ fixed point is not ruled out, are shown. $*$ This enhancement also requires two conserved currents that are flavor singlets with instanton number $\pm 2$. $**$ This enhancement also requires two conserved currents that are in the $(\bold{1},\bold{3})$ with instanton number $\pm 2$.} 
\label{summary2}
\end{table}

\subsubsection{SO(9)}

For $SO(9)$, there is a single spinor representation. The fermionic zero modes, provided by the spinors, are in the $(\bold{1},\bold{4})$ of the unbroken $SU(2)\times SO(5)$ gauge group. Next, we state our results for conserved currents.

We find only a handful of cases where there are conserved currents. First for $N_f=6, N_s=1$, there is a Higgs branch leading to $SU(2)+8F$ so this theory has a potential $6d$ origin. Recall that for $N_f=6, N_s=0$, there is a conserved current in the $\bold{12}$ of $USp_F(12)$. The additional fermionic zero modes charge it in the $\bold{3}$ of $SU_S(2)$ so we find a conserved current in the $(\bold{12},\bold{3})$ of $USp_F(12)\times SU_S(2)$. This cannot be contained in a finite Lie group, but rather in an affine one, $A^{(2)}_{14}$.

For $N_f=5$, we can have at most one spinor, as for $N_s=2$ there is a Higgs branch leading to two copies of $SU(2)+8F$. Again, recall that for $N_s=0$ there is a single conserved current which is a $USp_F(10)$ singlet. The additional spinor zero modes charge it under $SU_S(2)$, and we find a conserved current in the $(\bold{1},\bold{3})$ of $USp_F(10)\times SU_S(2)$. This suggests an enhancement of $U_T(1)\times SU_S(2)\rightarrow USp(4)$. For $ N_s=2$, it is now charged in the $(\bold{1},\bold{14})$ of $USp_F(10)\times USp_S(4)$. In addition we also find a conserved current in the $(\bold{44},\bold{1})$. These cannot be contained in a finite Lie group, rather suggesting both groups lift to affine Lie groups, particularly, $A^{(2)}_9$ and $A^{(2)}_4$.

For $N_f=4$ we find a conserved current only if $N_s\geq 2$. Specifically for $N_s=2$ we find a conserved current in the $(\bold{8},\bold{1})$ of $USp_F(8)\times USp_S(4)$. This should lead to an enhancement of $U_T(1)\times USp_F(8)\rightarrow USp(10)$ which should also require conserved currents with instanton number $\pm 2$ that are flavor symmetry singlets. This is the maximal allowed number of spinors as for $N_s=3$ there is a Higgs branch leading to $SU(2)+10F$. 

When $N_f=3$ we again find a conserved current only when $N_s\geq 2$. This current is a flavor singlet, and should lead to an enhancement of $U_T(1)\rightarrow SU(2)$. This is the maximal allowed number of spinors as for $N_s=3$ there is a Higgs branch leading to $SU(2)+8F$. Indeed, in that case, we find a conserved current in the $(\bold{1},\bold{21})$ of $USp_F(6)\times USp_S(6)$. Thus, $USp_S(6)$ appear to be affinized to $C^{(1)}_3$. Interestingly, $USp_F(6)$ is not affinized at this level.

The only other relevant case where we find a conserved current is $N_f=1, N_s=4$. Note that this theory has a Higgs branch leading to two copies of $SU(2)+8F$ so we do not expect a $5d$ fixed point, but a $6d$ one is possible. Indeed, we find a conserved current in the $(\bold{1},\bold{42})$ of $SU_F(2)\times USp_S(8)$ suggesting an enhanced affine Lie group $E^{(2)}_{6}$. The vector flavor symmetry is not affinized at this level.

We summarize our results for the conserved currents, and the expected symmetry enhancement, for theories with a potential $5d$ fixed point, in table \ref{summary3}.

\begin{table}[h!]
\begin{center}
\begin{tabular}{|c|c|c|c|}
  \hline 
   $N_f$ & $N_s$ & Currents & Minimal enhanced symmetry \\
 \hline
  $3$ & $2$ & $(\bold{1},\bold{1})$  & $USp_F(6)\times USp_S(4) \times SU(2)$ \\
 \hline
$4$ & $2$ & $(\bold{8},\bold{1})$ & $USp_S(4) \times USp(10)^{*}$ \\
 \hline
$5$ & $1$ & $(\bold{1},\bold{3})$ & $USp_F(10)\times USp(4)$  \\
 \hline
\end{tabular}
 \end{center}
\caption{The enhancement of symmetry for the $5d$ theory $SO(9)+N_f\bold{9}+N_s\bold{16}$. The classical global symmetry is $U_T(1)\times USp(2N_f)\times USp(2N_s)$. Written are the found conserved currents with their representation under the $USp(2N_f)\times USp(2N_s)$ global symmetry, and the minimal enhanced symmetry consistent with these currents. Only cases where conserved currents were found, and where a $5d$ fixed point is not ruled out, are shown. $*$ This enhancement also requires two conserved currents that are flavor singlets with instanton number $\pm 2$.} 
\label{summary3}
\end{table}

\subsubsection{SO(10)}

The spinor representation of $SO(10)$ is complex and the two chiralities are complex conjugates. The fermionic zero modes, provided by the spinors, are in the $(\bold{1},\bold{4})$ of the unbroken $SU(2)\times SO(6)$ gauge group. Next, we state our results for conserved currents.

For $N_f=7, N_s=1$ there is a Higgs branch leading to $SU(2)+8F$ so this theory may have a $6d$ origin. Indeed, we find two conserved currents with charges $\pm 2$ under $U_S(1)$ and in the $\bold{14}$ of $USp_F(14)$. This suggests an enhancement to the affine Lie group $A^{(2)}_{15}$. 

For $N_f=6$, the maximal allowed number of spinors is $N_s=1$. In that case, we find two conserved currents of charges $\pm 2$ under $U_S(1)$. This suggests an enhancement of $U_T(1)\times U_S(1)\rightarrow SU(2)^2$. The case of $N_s=2$ has a potential $6d$ origin, and we find several conserved currents with charges $(\bold{1},\bold{1})^{4}$, $(\bold{1},\bold{1})^{-4}$, $(\bold{1},\bold{5})^{0}$, and $(\bold{65},\bold{1})^{0}$ under $USp_F(12)\times SU_S(2)\times U_S(1)$. The last two suggest an enhancement of $USp_F(12)$ to the affine $A^{(2)}_{11}$ and of $SU_S(2)$ to $A^{(2)}_2$. In that light, we expect the first two currents to affinize $U_S(1)$, where the minimal possibility consistent with this data seems to be $A^{(1)}_1$.  

For $N_f=5$ we find a conserved current only if $N_s\geq 2$. A Higgs branch analysis reveals that the only relevant case is $N_s=2$ in which we find a conserved current which is in the $\bold{10}$ of $USp_F(10)$. This suggests an enhancement of $USp_F(10)\times U_T(1)\rightarrow USp(12)$ which also requires two flavor singlet currents with instanton charges $\pm 2$. 

For $N_f=4$, we again find conserved currents only if $N_s\geq 2$. Again, for a $5d$ fixed point, the only relevant case seems to be $N_s=2$ where we find a conserved current which is a flavor singlet. This suggests an enhancement of $U_T(1)\rightarrow SU(2)$. The case of $N_s=3$ is also interesting as it may have a $6d$ fixed point. In this case, we find two conserved currents of charges $\bold{6}^2$ and $\bar{\bold{6}}^{-2}$ under $SU_S(3)\times U_S(1)$. This suggests an affinization of $U_S(3)$ to $C^{(1)}_3$ while $USp_F(8)$ is unaffinized at this level. 

For $N_f=2$ we find conserved currents only if $N_s\geq 4$. Since for $N_s=4$ there is a Higgs branch leading to two copies of $SU(2)+8F$, this theory is the only relevant case. In that case, we find a conserved current in the $\bold{20'}$ of $SU_S(4)$ and two conserved currents with charges $\pm 4$ under $U_S(1)$ and in the $\bold{5}$ of $USp_F(4)$. The former seems to affinize $SU_S(4)$ to $A^{(2)}_5$ while the later seems to combine $U_S(1)\times USp_F(4)$ to the affine $B^{(1)}_3$.

For $N_f=1$, we again find conserved currents only if $N_s\geq 4$. With that number of spinors, there is always a Higgs branch leading to $SU(2)+(4N_s-9)F$ so the only relevant case is $N_s=4$ where we find two conserved currents of charges $\pm 4$ under $U_S(1)$ and in the $\bold{2}$ of $SU_F(2)$. The smallest global symmetry consistent with that is $SU(4)$, which also requires two flavor singlet currents with instanton charges $\pm 2$.

Finally, we consider the $N_f=0$ case. Again we find conserved currents only if $N_s\geq 4$. With that number of spinors, there is always a Higgs branch leading to $SU(2)+(4N_s-11)F$ so the only relevant case is $N_s=4$ where we find two conserved currents of charges $\pm 4$ under $U_S(1)$. This suggests an enhancement of $U_T(1)\times U_S(1)\rightarrow SU(2)^2$.

We summarize our results for the conserved currents, and the expected symmetry enhancement, for theories with a potential $5d$ fixed point, in table \ref{summary4}.

\begin{table}[h!]
\begin{center}
\begin{tabular}{|c|c|c|c|}
  \hline 
   $N_f$ & $N_s$ & Currents & Minimal enhanced symmetry \\
\hline
  $0$ & $4$ & $\bold{1}^{\pm 4}$ & $SU_S(4)\times SU(2)^2$ \\ 
\hline
  $1$ & $4$ & $(\bold{2},\bold{1})^{\pm 4}$ & $SU_S(4)\times SU(4)^{*}$ \\ 
\hline
  $4$ & $2$ & $(\bold{1},\bold{1})^0$  & $USp_F(8) \times SU_S(2) \times SU(2) \times U_S(1)$ \\
 \hline
$5$ & $2$ & $(\bold{10},\bold{1})^0$ & $U_S(1)\times SU_S(2) \times USp(12)^{*}$ \\
 \hline
$6$ & $1$ & $\bold{1}^{\pm 2}$ & $USp_F(12)\times SU(2)^2$  \\
 \hline
\end{tabular}
 \end{center}
\caption{The enhancement of symmetry for the $5d$ theory $SO(10)+N_f\bold{10}+N_s\bold{16}$. The classical global symmetry is $USp(2N_f)\times SU_S(N_s)\times U_S(1)\times U_T(1)$. Written are the found conserved currents with their representation under the $USp(2N_f)\times SU_S(N_s)\times U_S(1)$ global symmetry, and minimal enhanced symmetry consistent with these currents. Only cases where conserved currents were found, and where a $5d$ fixed point is not ruled out, are shown. $*$ This enhancement also requires two conserved currents that are flavor singlets with instanton number $\pm 2$.} 
\label{summary4}
\end{table}

\subsubsection{SO(11)}

There is a single spinor representation of $SO(11)$. This representation is pseudoreal and we can add half-hypers. The fermionic raising operators, provided by a spinor hyper, are in the $(\bold{1},\bold{8})$ of the unbroken $SU(2)\times SO(7)$ gauge group. A half-hyper in the spinor representation gives $8$ fermionic zero modes, in the $(\bold{1},\bold{8})$, that can be combined to form $4$ fermionic raising operators. Their application on the ground state generate $16$ states that decomposes to the $\bold{1} \oplus \bold{7} \oplus \bold{8}$ under the $SO(7)$ unbroken gauge group. Next, we state our results for conserved currents.

A Higgs branch analysis suggest that we cannot have more than $\frac{5}{2}$ spinors. The case of $\frac{5}{2}$ spinors cannot have a $5d$ fixed point though it may have a $6d$ one. Nevertheless, we find no conserved currents in this case. 

For $N_s=2$ We find a conserved current only if $N_f\geq 1$. When $N_f=1$ this conserved current is a flavor singlet, and is expected to lead to an enhancement of $U_T(1)\rightarrow SU(2)$. For $N_f=2$, the additional fermionic vector zero modes charge this current in the $\bold{4}$ of $USp_F(4)$. This suggests an enhancement of $USp_F(4)\times U_T(1)\rightarrow USp(6)$, which also requires two flavor singlet currents with instanton charges $\pm 2$. Finally, for $N_f=3$, there is a Higgs branch leading to $SU(2)+8F$ so this theory is still interesting as it may lift to $6d$. We find several conserved currents with charges $(\bold{1},\bold{1},\bold{14})$, $(\bold{1},\bold{5},\bold{1})$, and $(\bold{5},\bold{1},\bold{1})$ under $SU_{S_1}(2)\times SU_{S_2}(2) \times USp_F(6)$. This suggests an affinization of all the flavor symmetries where both $SU_S(2)$'s get lifted to $A^{(2)}_2$, and $USp_F(6)$ get lifted to $A^{(2)}_5$.

When $N_s=\frac{3}{2}$, we find conserved currents only when $N_f\geq 5$. Since for $N_f=5$ there is a Higgs branch leading to $SU(2)+8F$, this is the only interesting case. In that case, we find two conserved currents, one a flavor singlet and one in the $\bold{5}$ of $SU_S(2)$. The current in the $\bold{5}$ is consistent with $SU_S(2)$ getting lifted to the affine $A^{(2)}_2$, but the singlet does not appear to fit in this group. The simplest way to accommodate the singlet is by a twisted $U(1)$ group, that is a $U(1)$ symmetry group twisted by its charge conjugation outer automorphism. 

For $N_s=1$, we find conserved currents only when $N_f\geq 5$. When $N_f=5$, we find a conserved current which is a flavor singlet. This suggests an enhancement of $U_T(1)\rightarrow SU(2)$. For $N_f=6$ the additional vector fermionic zero modes furnish this current with the $\bold{12}$ dimensional representation of $USp_F(12)$. This suggests an enhancement of $USp_F(12)\times U_T(1)\rightarrow USp(14)$, which also requires two flavor singlet currents with instanton charges $\pm 2$. Finally, for $N_f=7$, there is a Higgs branch leading to two copies of $SU(2)+8F$ so this theory may have a $6d$ fixed point though not a $5d$ one. In that case, we find several conserved currents with charges $\bold{1}^{4}$, $\bold{1}^{-4}$, $\bold{1}^{0}$, and $\bold{90}^{0}$ under $USp_F(14)\times U_S(1)$. The last one suggests an enhancement of $USp_F(14)$ to the affine $A^{(2)}_{13}$. We expect the first two currents to also lift $U_S(1)$ to $6d$, the simplest option seems to be $A^{(1)}_1$. Finally the singlet can be accommodated by a twisted $U(1)$ group.

The last case to consider is $N_s=\frac{1}{2}$. We find that all the conserved currents come from the ground state which is a gauge $SO(7)$ singlet. Thus, the conserved current spectrum is identical to the case of $N_s=0$. More specifically, there are no conserved currents if $N_f<7$. For $N_f=7$ there is a conserved current which is a flavor singlet. This suggests an enhancement of $U_T(1)\rightarrow SU(2)$. Finally, for $N_f=8$ there is a conserved current which is in the $\bold{16}$ of $USp_F(16)$. This can be accommodated in a finite group, $USp(18)$, if in addition there are two flavor singlet currents with instanton charges $\pm 2$. However, this theory has a Higgs branch leading to $SU(2)+8F$ so we do not expect a $5d$ fixed point. 

We summarize our results for the conserved currents, and the expected symmetry enhancement, for theories with a potential $5d$ fixed point, in table \ref{summary5}.

\begin{table}[h!]
\begin{center}
\begin{tabular}{|c|c|c|c|}
  \hline 
   $N_f$ & $N_s$ & Currents & Minimal enhanced symmetry \\
\hline
  $1$ & $2$ & $(\bold{1},\bold{1})$  & $SU_F(2)\times SO_S(4) \times SU(2)$ \\ 
\hline
  $2$ & $2$ & $(\bold{4},\bold{1})$ & $SO_S(4) \times USp(6)^{*}$ \\ 
\hline
  $5$ & $1$ & $\bold{1}^0$  & $USp_F(10)\times SO_S(2) \times SU(2)$ \\
 \hline
$6$ & $1$ & $\bold{12}^0$ & $SO_S(2) \times USp(14)^{*}$ \\
 \hline
$7$ & $\frac{1}{2}$ & $\bold{1}$  & $USp_F(14)\times SU(2)$  \\
 \hline
\end{tabular}
 \end{center}
\caption{The enhancement of symmetry for the $5d$ theory $SO(11)+N_f\bold{11}+N_s\bold{32}$. The classical global symmetry is $USp(2N_f)\times SO_S(2N_s)\times U_T(1)$. Written are the found conserved currents with their representation under the $USp(2N_f)\times SO_S(2N_s)$ global symmetry, and minimal enhanced symmetry consistent with these currents. Only cases where conserved currents were found, and where a $5d$ fixed point is not ruled out, are shown. $*$ This enhancement also requires two conserved currents that are flavor singlets with instanton number $\pm 2$.} 
\label{summary5}
\end{table}

\subsubsection{SO(12)}

For $SO(12)$ there are two different, self conjugate, spinor representations. Both are pseudoreal so half-hypers in these representations are possible. The fermionic raising operators, provided by a spinor hyper, are in the $(\bold{1},\bold{8}_{S/C})$ of the unbroken $SU(2)\times SO(8)$ gauge group, depending on the chirality of the spinor. A half-hyper in the spinor representation gives $8$ fermionic zero modes, in the $(\bold{1},\bold{8}_{S/C})$, that can be combined to form $4$ fermionic raising operators. Their application on the ground state generates $16$ states that decomposes to the $ \bold{8}_V \oplus \bold{8}_{C/S}$ under the $SO(8)$ unbroken gauge group. Next, we state our results for conserved currents.

Like for $SO(11)$, the maximal number of spinor half-hypers, in any combination of chirality, is $\frac{5}{2}$. When all of these are of the same chirality, the theory has a Higgs branch leading to the maximally supersymmetric $SU(2)$ gauge theory so this theory may have a $6d$ fixed point, but probably not a $5d$ one. Nevertheless, we find no conserved currents in this case.

 The other possibilities, $N_s=2, N_c=\frac{1}{2}$ and  $N_s=\frac{3}{2}, N_c=1$, with $N_f$ vectors, have a Higgs branch leading to $USp(6)+\frac{1}{2}TAS+1AS+(2N_f+\frac{1}{2})F$ and $USp(6)+1TAS+(2N_f+3)F$ respectively. From the analysis of section $3.3$, we conclude that the case of $N_f=0$ may have a $5d$ fixed point while the $N_f=1$ case may have a $6d$ one. 

For the $N_f=0, N_s=2, N_c=\frac{1}{2}$ case, we find a conserved current which is in the $(\bold{2},\bold{2})$ of $SU_S(2)^2$. This suggests an enhancement of $SU_S(2)^2\times U_T(1)\rightarrow SU(4)$. For the $N_f=0, N_s=\frac{3}{2}, N_c=1$ case, we find two conserved currents with charges $\pm 3$ under $U_C(1)$. This suggests an enhancement of $U_C(1)\times U_T(1)\rightarrow SU(2)^2$. 

Finally, for $N_f=1, N_s=2, N_c=\frac{1}{2}$, we find a conserved current which is in the $(\bold{2},\bold{2},\bold{2})$ of $SU_S(2)^2\times SU_F(2)$. This can fit in a finite group leading to an enhancement of $SU_S(2)^2\times SU_F(2)\times U_T(1)\rightarrow SO(8)$, assuming there are additional flavor singlet currents with instanton charges $\pm 2$. For $N_f=1, N_s=\frac{3}{2}, N_c=1$, we find two conserved currents in the $\bold{2}$ of $SU_F(2)$ and with charge $\pm 3$ under $U_C(1)$. This can fit in a finite group leading to an enhancement of $ SU_F(2)\times U_C(1) \times U_T(1)\rightarrow SU(4)$, assuming there are additional flavor singlet currents with instanton charges $\pm 2$.

Next we discuss the case of $4$ spinor half-hypers. First consider the case where all of them are of the same chirality, and $N_f=0$. In that case, we find a conserved current which is a flavor singlet. This suggests an enhancement of $U_T(1)\rightarrow SU(2)$. 

Adding vectors will now furnish this current in the rank $N_f$ antisymmetric irreducible representation of $USp_F(2N_f)$. So for $N_f=1$, we find a conserved current in the $\bold{2}$ of $SU_F(2)$. This suggests an enhancement of $U_T(1)\times SU_F(2)\rightarrow SU(3)$. For $N_f=2$, the current is now in the $\bold{5}$ of $USp_F(4)$, and we expect  an enhancement of $U_T(1)\times USp_F(4)\rightarrow SO(7)$. Finally, for $N_f=3$ the conserved current is in the $\bold{14}'$ of $USp_F(6)$. This suggests an enhancement of $U_T(1)\times USp_F(6)\rightarrow F_4$ which also requires two flavor singlet currents with instanton charges $\pm 2$.

For $N_f=4$, there is a Higgs branch leading to $SU(2)+8F$, and indeed the conserved current is now in the $\bold{42}$ of $USp_F(8)$. This cannot be accommodated in a finite Lie group, but rather in the affine $E^{(2)}_6$, as could be expected from a theory with a $6d$ fixed point. Moreover, we find two additional currents with charges $(\bold{1},\bold{5},\bold{1})$, and $(\bold{5},\bold{1},\bold{1})$ under $SU_{S_1}(2)\times SU_{S_2}(2) \times USp_F(8)$. This suggests that both $SU_S(2)$'s also get lifted to $A^{(2)}_2$.

If instead we consider $N_s=\frac{3}{2}, N_c=\frac{1}{2}$ then we find no conserved currents unless $N_f\geq 4$. As for $N_f=4$ there is a Higgs branch leading to $SU(2)+8F$, this is the only interesting case. In this case, we find a conserved current in the $\bold{5}$ of $SU_S(2)$. This suggests an affinization of $SU_S(2)$ to $A^{(2)}_2$ while $USp_F(8)$ appears not to affinize at this level.

The last case to consider is $N_s=N_c=1$. In this case we find no conserved currents unless $N_f\geq 2$. For $N_f=2$ this current is a flavor singlet, and we expect an enhancement of $U_T(1)\rightarrow SU(2)$. For $N_f=3$, the additional vector fermionic zero modes charge the current in the $\bold{6}$ of $USp_F(6)$. This suggests an enhancement of $USp_F(6)\times U_T(1)\rightarrow USp(8)$ which also requires two flavor singlet currents with instanton charges $\pm 2$.

 For $N_f=4$, there is a Higgs branch leading to $SU(2)+8F$ so this case is again interesting from a $6d$ perspective. We find several conserved currents with charges $\bold{1}^{(0,0)}$, $\bold{1}^{(2,2)}$, $\bold{1}^{(2,-2)}$, $\bold{1}^{(-2,2)}$, $\bold{1}^{(-2,-2)}$, and $\bold{27}^{(0,0)}$ under $USp_F(8)\times U_S(1)\times U_C(1)$. The last one suggests an affinization of $USp_F(8)$ to $A^{(2)}_{7}$. Again, in that light we expect the remaining currents to affinize $U_S(1)$ and $U_C(1)$, where the simplest option seems to be $A^{(1)}_1 \times A^{(1)}_1$. The singlet can be accommodated by a twisted $U(1)$ group.

We next consider the case of three half-hypers. From all the relevant cases, we find only one where there is a conserved current. This case is $N_f=6, N_s=1, N_c=\frac{1}{2}$, and we find two conserved currents with charges $\pm 1$ under $U_S(1)$. These can fit in a finite group, particularly, $U_T(1)\times U_S(1)\rightarrow SU(2)^2$. However, this theory has a Higgs branch leading to $SU(2)+8F$ so we do not expect a $5d$ fixed point. 

Next we discuss the case where there are two half-hypers. There are just two possibilities, either the two half-hypers are of the same chirality or opposite ones. In either case, we find a conserved current only when $N_f\geq 6$. Specifically, for the $N_f=6$ case, we find a conserved current which is a flavor singlet in both cases. This suggests an enhancement of $U_T(1)\rightarrow SU(2)$. When $N_f=7$, this current is charged in the $\bold{14}$ of $USp_F(14)$. So in both cases, we expect an enhancement of $USp_F(14)\times U_T(1)\rightarrow USp(16)$ which also requires two flavor singlet currents with instanton charges $\pm 2$.

Finally, for $N_f=8$, there is a Higgs branch leading to two copies of $SU(2)+8F$ so this theory probably doesn't have a $5d$ fixed point, but may have a $6d$ one. In this case, we find different currents in both cases. In the case of $N_s=1, N_c=0$, we find several conserved currents with charges $\bold{1}^{4}$, $\bold{1}^{-4}$, and $\bold{119}^{0}$ under $USp_F(16)\times U_S(1)$. The last one suggests an enhancement of $USp_F(16)$ to the affine $A^{(2)}_{15}$. We expect the first two to affinize $U_S(1)$, where the simplest option seems to be $A^{(1)}_1$. 

For $N_s=\frac{1}{2}, N_c=\frac{1}{2}$, we find two conserved current one a flavor singlet while the other is in the $\bold{119}$ of $USp_F(16)$. The last one suggests an enhancement of $USp_F(16)$ to the affine $A^{(2)}_{15}$, and a twisted $U(1)$ group coming from the singlet.

Finally, there is the case of a single half-hyper. Going over all the relevant cases, we find no conserved current. We summarize our results for the conserved currents, and the expected symmetry enhancement, for theories with a potential $5d$ fixed point, in table \ref{summary6}.

\begin{table}[h!]
\begin{center}
\begin{tabular}{|c|c|c|c|c|}
  \hline 
   $N_f$ & $N_s$ & $N_c$ & Currents & Minimal enhanced symmetry \\
\hline
  $0$ & $2$ &  $\frac{1}{2}$ & $\bold{4}_V$  & $SU(4)$ \\
\hline
  $0$ & $\frac{3}{2}$ & $1$ & $\bold{1}^{\pm 3}$ & $SO_S(3)\times SU(2)^2$ \\
\hline
  $0$ & $2$ & $0$ & $\bold{1}$  & $SO_S(4)\times SU(2)$ \\ 
\hline
  $1$ & $2$ & $0$ & $(\bold{2},\bold{1})$  & $SO_S(4)\times SU(3)$ \\ 
\hline
  $2$ & $2$ & $0$ & $(\bold{5},\bold{1})$ & $SO_S(4)\times SO(7)$ \\
\hline
  $3$ & $2$ & $0$ & $(\bold{14}',\bold{1})$ & $SO_S(4)\times F_4 ^{*}$ \\
\hline
  $2$ & $1$ & $1$ & $\bold{1}^{(0,0)}$  & $USp_F(4)\times SO_S(2) \times SO_C(2) \times SU(2)$ \\
 \hline
$3$ & $1$ & $1$ & $\bold{6}^{(0,0)}$ & $SO_S(2) \times SO_C(2) \times USp(8)^{*}$ \\
 \hline
$6$ & $1$ & $0$ & $\bold{1}^0$  & $USp_F(12)\times SO_S(2) \times SU(2)$ \\
 \hline
$6$ & $\frac{1}{2}$ & $\frac{1}{2}$ & $\bold{1}$  & $USp_F(12)\times SU(2)$ \\
 \hline
$7$ & $1$ & $0$ & $\bold{14}^0$ & $SO_S(2) \times USp(16)^{*}$ \\
 \hline
$7$ & $\frac{1}{2}$ & $\frac{1}{2}$ & $\bold{14}$ & $USp(16)^{*}$ \\
 \hline
\end{tabular}
 \end{center}
\caption{The enhancement of symmetry for the $5d$ theory $SO(12)+N_f\bold{12}+N_s\bold{32}+N_c\bold{32}'$. The classical global symmetry is $USp(2N_f)\times SO_S(2N_s)\times SO_C(2N_c)\times U_T(1)$. Written are the found conserved currents with their representation under the $USp(2N_f)\times SO_S(2N_s)\times SO_C(2N_c)$ global symmetry, and minimal enhanced symmetry consistent with these currents. Only cases where conserved currents were found, and where a $5d$ fixed point is not ruled out, are shown. $*$ This enhancement also requires two conserved currents that are flavor singlets with instanton number $\pm 2$.} 
\label{summary6}
\end{table}

\subsubsection{SO(13)}

For $SO(13)$ there is a single spinor representation. This representation is pseudoreal so half-hypers in this representation are possible. The fermionic raising operators, provided by a spinor hyper, are in the $(\bold{1},\bold{16})$ of the unbroken $SU(2)\times SO(9)$ gauge group. A half-hyper in the spinor representation gives $16$ fermionic zero modes, in the $(\bold{1},\bold{16})$, that can be combined to form $8$ fermionic raising operators. Their application on the ground state generate $256$ states that decomposes to the $\bold{44} \oplus \bold{84} \oplus \bold{128}$ under the $SO(9)$ unbroken gauge group. Next, we state our results for conserved currents.

A Higgs branch analysis suggests that we cannot have more than two spinor half-hypers. For $N_s=1$, we find conserved currents only for $N_f \geq 3$. For $N_f=3$, this current is a flavor singlet, and we expect an enhancement of $U_T(1)\rightarrow SU(2)$. For $N_f=4$, this current acquires the representation $\bold{8}$ under $USp_F(8)$. This suggests an enhancement of $USp_F(8)\times U_T(1)\rightarrow USp(10)$, which also requires two flavor singlet currents with instanton charges $\pm 2$. 

Finally, for $N_f=5$, there is a Higgs branch leading to $SU(2)+8F$ so this theory may have a $6d$ fixed point. In that case, we find several conserved currents with charges $\bold{1}^{4}$, $\bold{1}^{-4}$, $\bold{1}^{0}$, and $\bold{44}^{0}$ under $USp_F(10)\times U_S(1)$. The last one suggests an enhancement of $USp_F(10)$ to the affine $A^{(2)}_{9}$. We expect the first two to also affinize $U_S(1)$, where the simplest option seems to be $A^{(1)}_1$. The singlet can be accommodated by a twisted $U(1)$ group.  

In the case of $N_s=\frac{1}{2}$, we find conserved currents only when $N_f \geq 7$. For $N_f=7$, this conserved current is a flavor singlet. This suggests an enhancement of $U_T(1)\rightarrow SU(2)$. For $N_f=8$, this current is now in the $\bold{16}$ of $USp_F(16)$. This suggests an enhancement of $USp_F(16)\times U_T(1)\rightarrow USp(18)$ which also requires two flavor singlet currents with instanton charges $\pm 2$. Finally, for $N_f=9$, we find a conserved current in the $\bold{152}$ of $USp_F(18)$. This does not fit in a finite Lie group, but rather in the affine $A^{(2)}_{17}$. This is consistent with the Higgs branch analysis where this theory reduces to two copies of $SU(2)+8F$.

We summarize our results for the conserved currents, and the expected symmetry enhancement, for theories with a potential $5d$ fixed point, in table \ref{summary7}.

\begin{table}[h!]
\begin{center}
\begin{tabular}{|c|c|c|c|}
  \hline 
   $N_f$ & $N_s$ & Currents & Minimal enhanced symmetry \\
\hline
  $3$ & $1$ & $\bold{1}^0$  & $USp_F(6)\times SO_S(2)\times SU(2)$ \\ 
\hline
  $4$ & $1$ & $\bold{8}^0$ & $SO_S(2)\times USp(10)^{*}$ \\ 
\hline
  $7$ & $\frac{1}{2}$ & $\bold{1}$  & $USp_F(14)\times SU(2)$ \\
 \hline
$8$ & $\frac{1}{2}$ & $\bold{16}$ & $USp(18)^{*}$ \\
 \hline
\end{tabular}
 \end{center}
\caption{The enhancement of symmetry for the $5d$ theory $SO(13)+N_f\bold{13}+N_s\bold{64}$. The classical global symmetry is $USp(2N_f)\times SO_S(2N_s)\times U_T(1)$. Written are the found conserved currents with their representation under the $USp(2N_f)\times SO_S(2N_s)$ global symmetry, and minimal enhanced symmetry consistent with these currents. Only cases where conserved currents were found, and where a $5d$ fixed point is not ruled out, are shown. $*$ This enhancement also requires two conserved currents that are flavor singlets with instanton number $\pm 2$.} 
\label{summary7}
\end{table}  

%- $N_f=7, N_s=\frac{1}{2}$: We find a conserved current which is a flavor singlet. This suggests an enhancement of $U_T(1)\rightarrow SU(2)$.

%- $N_f=8, N_s=\frac{1}{2}$: We find a conserved current which is in the $\bold{16}$ of $USp_F(16)$. This suggests an enhancement of $USp_F(16)\times U_T(1)\rightarrow USp(18)$ which also requires two flavor singlet currents with instanton charges $\pm 2$.

%- $N_f=3, N_s=1$: We find a conserved current which is a flavor singlet. This suggests an enhancement of $U_T(1)\rightarrow SU(2)$.

%- $N_f=4, N_s=1$: We find a conserved current which is in the $\bold{8}$ of $USp_F(8)$. This suggests an enhancement of $USp_F(8)\times U_T(1)\rightarrow USp(10)$ which also requires two flavor singlet currents with instanton charges $\pm 2$.

%For theories with a potential $6d$ origin, we find:

%- $N_f=9, N_s=\frac{1}{2}$: We find a conserved current in the $\bold{152}$ of $USp_F(18)$. This does not fit in a finite Lie group, but rather in the affine $A^{(2)}_{17}$

%- $N_f=5, N_s=1$: We find several conserved current with charges $\bold{1}^{4}$, $\bold{1}^{-4}$, $\bold{1}^{0}$, and $\bold{44}^{0}$ under $USp_F(10)^{U_S(1)}$. The last one suggests an enhancement of $USp_F(10)$ to the affine $A^{(2)}_{9}$.

\subsubsection{SO(14)}

For $SO(14)$ there are two complex spinors where each one is the complex conjugate of the other. The fermionic raising operators, provided by a spinor hyper, are in the $(\bold{1},\bold{16})$ of the unbroken $SU(2)\times SO(10)$ gauge group. Next, we state our results for conserved currents.

A Higgs branch analysis suggests we must have $ N_s \leq 1$ so the only relevant case is $N_s=1$. The same analysis also suggests that we must have $N_f \leq 5$. We find conserved currents only when $N_f \geq 4$. When $N_f=4$ this current is a flavor singlet, and we expect an enhancement of $U_T(1)\rightarrow SU(2)$. When $N_f=5$, this current is in the $\bold{10}$ of $USp_F(10)$. This suggests an enhancement of $USp_F(10)\times U_T(1)\rightarrow USp(12)$, which also requires two flavor singlet currents with instanton charges $\pm 2$. 

Finally, for $N_f=6$, there is a Higgs branch leading to $SU(2)+8F$ so this case is still interesting as there may be a $6d$ fixed point. In this case, we find several conserved currents with charges $\bold{1}^{4}$, $\bold{1}^{-4}$, and $\bold{65}^{0}$ under $USp_F(12)\times U_S(1)$. The last one suggests an enhancement of $USp_F(12)$ to the affine $A^{(2)}_{11}$, and we expect the first two to also affinize $U_S(1)$, where the simplest option seems to be $A^{(1)}_1$.  

We summarize our results for the conserved currents, and the expected symmetry enhancement, for theories with a potential $5d$ fixed point, in table \ref{summary8}.

\begin{table}[h!]
\begin{center}
\begin{tabular}{|c|c|c|c|}
  \hline 
   $N_f$ & $N_s$ & Currents & Minimal enhanced symmetry \\
\hline
  $4$ & $1$ & $\bold{1}^0$  & $USp_F(8)\times SO_S(2) \times SU(2)$ \\ 
\hline
  $5$ & $1$ & $\bold{10}^0$ & $SO_S(2) \times USp(12)^{*}$ \\ 
\hline
\end{tabular}
 \end{center}
\caption{The enhancement of symmetry for the $5d$ theory $SO(14)+N_f\bold{14}+N_s\bold{64}$. The classical global symmetry is $USp_F(2N_f)\times SU_S(N_s)\times U_S(1)\times U_T(1)$. Written are the found conserved currents with their representation under the $USp_F(2N_f)\times SU_S(N_s)\times U_S(1)$ global symmetry, and minimal enhanced symmetry consistent with these currents. Only cases where conserved currents were found, and where a $5d$ fixed point is not ruled out, are shown. $*$ This enhancement also requires two conserved currents that are flavor singlets with instanton number $\pm 2$.} 
\label{summary8}
\end{table}  

%- $N_f=4, N_s=1$: We find a conserved current which is a flavor singlet. This suggests an enhancement of $U_T(1)\rightarrow SU(2)$.

%- $N_f=5, N_s=1$: We find a conserved current which is in the $\bold{10}$ of $USp_F(10)$. This suggests an enhancement of $USp_F(10)\times U_T(1)\rightarrow USp(12)$ which also requires two flavor singlet currents with instanton charges $\pm 2$.

%For theories with a potential $6d$ origin, we find:

%- $N_f=6, N_s=1$: We find several conserved current with charges $\bold{1}^{4}$, $\bold{1}^{-4}$, and $\bold{65}^{0}$ under $USp_F(12)^{U_S(1)}$. The last one suggests an enhancement of $USp_F(12)$ to the affine $A^{(2)}_{11}$.

\section{The case of exceptional groups}

In this section we discuss the case of exceptional groups. Like the case of $SO$ groups with spinors, this is especially interesting as conventional instanton counting methods are unavailable in this case. In the pure YM case this can be circumvented owing to the identification of the Nekrasov partition function\cite{NS} with the properly symmetrized Hilbert series of the corresponding instanton moduli space\cite{RG}. Fortunately, methods for the calculation of the Hilbert series for the instanton moduli spaces of exceptional groups are known\cite{BHM}. Using their results we find that the $1$-instanton Nekrasov partition function has the following form:

\be
Z^1_G = \frac{x^{h_G}}{(1-x y)(1-\frac{x}{y})} \sum^{\infty}_{k=0} \chi[k] x^{2k}
\ee 
where $h_G$ is the dual Coexter number of the group $G$ and $\chi[k]$ stands for the character of the irreducible $k$-symmetric product of the Adjoint (in Cartan weights these are given by $[0,k]$ for $G_2$, $[k,0,0,0]$ for $F_4$, $[0,k,0,0,0,0]$ for $E_6$, $[k,0,0,0,0,0,0]$ for $E_7$ and $[0,0,0,0,0,0,0,k]$ for $E_8$). This result has the following simple interpretation. The infinite series and the denominator come from applying the $4h_G$ bosonic zero modes on a single instanton ground state. This state in turn contributes $x^{h_G}$ to the index.

Next, we discus each group in turn, stating our results.

\subsection{$G_2$}

Instantons of $G_2$ are contained in an $SU(2)$ subgroup breaking $G_2\rightarrow SU(2)\times SU(2)$. Under this breaking the Adjoint $\bold{14}$ of $G_2$ decomposes into the Adjoints of both $SU(2)$'s as well as a state in the $(\bold{2},\bold{4})$ representation. Thus, the full state space is given by acting on the ground state, whose r-charge is $-1$, with the fermionic operator $B_i$ in the $\bold{4}$ of the unbroken $SU(2)$ gauge group. 

We can form an $SU(2)$ gauge invariant by contracting two $B_i$ operators. By applying these on the ground state we get a triplet corresponding to one state with R-charge $1$. Tensoring with the broken current supermultiplet, we find a single state which, while not a conserved current, is a BPS state. We expect this state to contribute the term $x^4$ to the index. Recalling that $h_{G_2}=4$, this matches our expectations from the partition function.

Next, we generalize by adding hypermultiplets in the $\bold{7}$ of $G_2$. Each one of these contributes a raising operator in the $\bold{2}$ of the unbroken $SU(2)$ gauge group. These can form an invariant by contraction with $\epsilon^{ij}$, whose application on the previous state furnishes it with the rank $N_f$ irreducible antisymmetric representation of the $USp(2N_f)$ flavor symmetry. 

Next we ask whether one can form a conserved current state by contracting both types of zero modes. In order to get a conserved current we need an R-charge singlet, necessitating two $B_i$'s. As these are fermionic zero modes, the $SU(2)$ charges must form an antisymmetric product so the product can be either in the singlet or the $\bold{5}$ dimensional representation. The singlet is just the previous state so if we are to find a conserved current it must come by contracting the $B_i$'s in the $\bold{5}$ dimensional representation of $SU(2)$ and canceling it against flavor zero modes. 

To get the $\bold{5}$ dimensional representation we need a symmetric product of $4$ states in the $\bold{2}$, and since these are fermionic zero modes, they must be antisymmetric in their flavor charges. Therefore, we conclude that for $N_f<4$ there are no conserved currents. When $N_f=4$ there is one conserved current which is a flavor symmetry singlet. This should lead to the enhancement $U_T(1)\rightarrow SU(2)$. 

When $N_f=5$ there is a conserved current which is in the $\bold{10}$ of the $USp_F(10)$ flavor symmetry. This should lead to the enhancement $U_T(1)\times USp_F(10) \rightarrow USp(12)$. This also requires conserved currents with instanton number $\pm 2$ that are flavor symmetry singlets. Finally, when $N_f=6$ the conserved current carries a rank $2$ irreducible antisymmetric tensor of the $USp_F(12)$ flavor symmetry. This cannot form a finite Lie group, but rather seems to lead to the affine $A^{(2)}_{11}$. This suggests that this theory lifts to $6d$.   

We can also consider the maximally supersymmetric case where we add a hypermultiplet in the $\bold{14}$ of $G_2$. The fermionic zero modes supplied by the Adjoint hyper form $4$ raising operators which together with $B_i$ form a doublet of $SU(2)\subset SO_R(5)$. The basic gauge invariant one can form is given by contracting two such operators. This results in a gauge invariant raising operator that is in the $\bold{3}$ of $SU(2)\subset SO_R(5)$. Acting on the ground state, with r-charge $-2$, with these operators generates $14$ states that form the rank $2$ symmetric traceless representation of $SO_R(5)$.

 The general thought is that this theory lift to the $6d$ $(2,0)$ theory of type $D_4$ where a $Z_3$ twist in the outer automorphism of $D_4$ is imposed when going around the circle\cite{Tachi1}. Recall that $SO(8)$ has invariant polynomials of degree $2,6$ and two distinct ones of degree $4$. Under the $Z_3$ twist the polynomials of degree $2,6$ are even, and so contribute at the massless level. This correctly reproduces the expected operators of $G_2$ whose invariant polynomials are of degrees $2$ and $6$. 

On the other hand, the two polynomials of degree $4$ are rotated by $120^0$\cite{Tachi2}. Thus, these states are all massive with mass $\sim \frac{1}{g^2_{YM}}$ where the lightest states having a mass of $\frac{M_{KK}}{3}, \frac{2M_{KK}}{3}$ where $M_{KK}$ is the mass of the first KK state given by the states corresponding to the invariant polynomials even under the twist. The first instanton contribution should correspond to the state with mass $\frac{M_{KK}}{3}$ and so we expect it to give a KK supermultiplet in the $\bold{14}$ of $SO_R(5)$. This indeed matches our results.   

\subsection{$F_4$}

Instantons of $F_4$ are contained in an $SU(2)$ subgroup breaking $F_4\rightarrow SU(2)\times USp(6)$. Under this breaking the Adjoint $\bold{52}$ of $F_4$ decomposes into the Adjoints of $SU(2)$ and $USp(6)$ as well as a state in the $(\bold{2},\bold{14}')$ representation. Thus, the full state space is given by acting on the ground state, whose r-charge is $-\frac{7}{2}$, with the fermionic operator $B_i$ in the $\bold{14}'$ of the unbroken $USp(6)$ gauge group. We can form a $USp(6)$ gauge invariant by contracting two $B_i$ operators. By applying these on the ground state we get a tower corresponding to one state with R-charge $\frac{7}{2}$. Tensoring this with the basic broken current supermultiplet, we get a BPS state, which we expect to contribute to the index as $x^9$. Recalling that $h_{F_4}=9$, this matches the expectations from the partition function.

We can generalize by adding hypermultiplets in the $\bold{26}$ of $F_4$. Each one of these contributes a raising operator in the $\bold{6}$ of the unbroken $USp(6)$ gauge group. We can form gauge invariants by contracting two such operators. Their application on the ground state will charge it under the flavor symmetry. We can ask whether we can form an invariant which will be an R-charge singlet so that we get a conserved current. Using numerical analysis, we find that this is only possible if $N_f>2$. For $N_f=3$, the conserved current is a flavor singlet suggesting an enhancement of $U_T(1)\rightarrow SU(2)$.

Incidentally, $N_f=3$ is the maximal possible number of flavors, since beyond this, one can flow on the Higgs branch to theories that are thought to possess neither a $5d$ nor a $6d$ fixed point. Indeed, solving numerically for $N_f=4$, we find a conserved current in the $\bold{120}$ of $USp_F(8)$. This cannot fit in either a finite or an affine Lie group. 

We can also consider the maximally supersymmetric case. Evaluating numerically, we find two distinct states in the rank $7$ and $3$ symmetric traceless representation of $SO_R(5)$. We expect this theory to lift to the $6d$ $(2,0)$ theory of type $E_6$. The reduction to $5d$ is done with a $Z_2$ outer automorphism twist under which the $E_6$ invariant polynomials of degrees $2,6,8$ and $12$ are even, while those of degrees $5$ and $9$ are odd. Therefore the first instanton contribution should be the lowest mode of the states associated to the invariant polynomials of degrees $5$ and $9$, and so should be in the states given by tensoring the KK state with the rank $3$ and $7$ symmetric traceless representation of $SO_R(5)$, respectively. This indeed matches our results.  

\subsection{$E_6$}

Instantons of $E_6$ are contained in an $SU(2)$ subgroup breaking $E_6\rightarrow SU(2)\times SU(6)$. Under this breaking the Adjoint $\bold{78}$ of $E_6$ decomposes into the Adjoints of $SU(2)$ and $SU(6)$ as well as a state in the $(\bold{2},\bold{20})$ representation. Thus, the full state space is given by acting on the ground state, whose r-charge is $-5$, with the fermionic operator $B_i$ in the $\bold{20}$ of the unbroken $SU(6)$ gauge group. We can form an $SU(6)$ gauge invariant by contracting two $B_i$ operators. By applying these on the ground state we get a tower corresponding to one state with R-charge $5$. This is a BPS state which we expect to contribute to the index as $x^{12}$. Recalling that $h_{E_6}=12$, this matches the expectations from the partition function.

We can generalize by adding hypermultiplets in the $\bold{27}$ of $E_6$. Each one of these contribute a raising operator in the $\bold{6}$ of the unbroken $SU(6)$ gauge group. We can ask whether we can form an invariant which will be an R-charge singlet so that we get a conserved current. Using numerical analysis, we find that this is only possible if $N_f=4$ in which we find a single conserved current which is a flavor singlet. This leads us to expect an enhancement of $U_T(1)\rightarrow SU(2)$ in this case. We do not expect a fixed point, $5d$ or $6d$, to exist when $N_f>4$.

We can also consider the maximally supersymmetric case. Unfortunately, the numerics in this case, as well as for the other $E$ groups, proves to be quite time consuming, so we reserve this for future study. 

\subsection{$E_7$}

Instantons of $E_7$ are contained in an $SU(2)$ subgroup breaking $E_7\rightarrow SU(2)\times SO(12)$. Under this breaking the Adjoint $\bold{133}$ of $E_7$ decomposes into the Adjoints of $SU(2)$ and $SO(12)$ as well as a state in the $(\bold{2},\bold{32})$ representation. Thus, the full state space is given by acting on the ground state, whose r-charge is $-8$, with the fermionic operator $B_i$ in the $\bold{32}$ of the unbroken $SO(12)$ gauge group. we can form an $SO(12)$ gauge invariant by contracting two $B_i$ operators. By applying these on the ground state we get a tower corresponding to one state with R-charge $8$. Taking the direct product with the broken current supermultiplet, we find a single state with R-charge $9$. Recalling that $h_{E_7}=18$, this matches our expectations from the partition function.

We can generalize by adding hypermultiplets in the $\bold{56}$ of $E_7$. This representation is pseudoreal so half-hypers are possible. From the Higgs branch analysis, we conclude that the maximal number of half-hypers possible for the existence of a $5d$ fixed point is $6$. Each full hyper contributes a raising operator in the $\bold{12}$ of the unbroken $SO(12)$ gauge group. A half-hyper contributes $12$ fermionic zero modes which can be combined to form $6$ raising operators in the $\bold{6}^1$ of $U(1)\times SU(6)\subset SO(12)$. Applying them on the ground state yields $64$ states transforming as the $\bold{32}\oplus \bold{32}'$ of $SO(12)$.

We now ask, for a given number of half-hypers, can we form a gauge invariant state with $0$ R-charge, which as we are now quite accustomed to gives a conserved current after tensoring with the broken current supermultiplet. From group theory it is clear that this can only happen when the number of half-hypers is even. First recall that $SO(12)$ has a $Z_2 \times Z_2$ center. This contains three distinct $Z_2$ subgroups, each of which act on two of the basic representations: $\bold{12}$, $\bold{32}$, and $\bold{32}'$, as $-1$ and on the other by $1$. 

If a state is to be gauge invariant, it must also be invariant under the center. The previous analysis now implies that only a product of an even number of all three basic representations, or a product of an odd number of all of them, can contain a gauge invariant. In particular, when applied to the case of an odd number of half-hypers, we see that an $SO(12)$ invariant state can only be made by an odd number of $B_i$'s and thus cannot be an R-charge singlet.

This leaves only $3$ cases that need to be worked out exactly. Unfortunately, the numerical analysis proves to be quite time consuming and we leave pursuing it to future work. 

\subsection{$E_8$}

Instantons of $E_8$ are contained in an $SU(2)$ subgroup breaking $E_8\rightarrow SU(2)\times E_7$. Under this breaking the Adjoint $\bold{248}$ of $E_8$ decomposes into the Adjoints of $SU(2)$ and $E_7$ as well as a state in the $(\bold{2},\bold{56})$ representation. Thus, the full state space is given by acting on the ground state, whose r-charge is $-14$, with the fermionic operator $B_i$ in the $\bold{56}$ of the unbroken $E_7$ gauge group. we can form an $E_7$ gauge invariant by contracting two $B_i$ operators. By applying these on the ground we get a tower corresponding to one state with R-charge $14$. When tensored with the broken current supermultiplet, this results in a BPS state which we expect to contribute a term of $x^{15}$ to the index. Recalling that $h_{E_8}=30$, this matches our expectations from the partition function.

The fundamental $\bold{248}$ representation of $E_8$ is identical to the Adjoint so upon adding a single hyper we get the maximally supersymmetric $E_8$ Yang-Mills theory, which should lift to $6d$. 

\section{Conclusions}

In this paper we explored symmetry enhancement from $1$ instanton operators for $USp(2N)$, $SO(N)$ and exceptional groups with various matter content. This line of thought can be generalized in several directions. First, one can attempt to explore more general $1$ instanton operators, where one applies also bosonic zero modes or fermionic non-zero modes, or go to higher instanton number. We have seen that in some cases, in order to complete a symmetry group, conserved currents with higher instanton number are necessary. It will be interesting if these can also be verified by some modification of these methods. 

We have also concentrated on the case of a simple gauge group. An interesting direction is to generalize to quiver theories, extending the results of \cite{Tachi}. Also, we have adopted a rather broad criterion, and it is not clear if the theories checked actually have a $5d$ fixed point. In some cases it is known to exist as these theories can be engineered in string theory using brane webs\cite{BZ1}. It is interesting to know whether the other cases also flow to $5d$ fixed points.

In some cases we have seen that the global symmetry is enhanced to an affine Lie group. The most straightforward interpretation of this is that these theories lift to a $6d$ theory with the appropriate global symmetry. In the case of $USp(2N)+AS+8F$ the $6d$ UV theory is known. It is interesting to see if we can understand the $6d$ lift also in the other cases.   

Recently, an explicit construction and study of instanton operators was initiated in \cite{LPS1,RS}. It is interesting if these constructions can also be generalized to the cases presented in this paper, and help shed light on the properties of these theories.

\subsection*{Acknowledgments}

I would like to thank Oren Bergman for useful comments and discussions. G.Z. is supported in part by the Israel Science Foundation under grant no. 352/13, and by the German-Israeli Foundation for Scientific Research and Development under grant no. 1156-124.7/2011.

\end{document}